\documentclass[preprint,times]{aastex63}

\usepackage{color}
\usepackage{natbib}
\usepackage{amsmath}
\usepackage{grffile}    % to allow periods in graphics filenames
\tightenlines

\newcommand \bomega     {{\boldsymbol{\omega}}}

\newcommand \bomegahat  {\hat{\boldsymbol{\omega}}}

\newcommand \Angstrom   {\,{\rm \AA}}

\newcommand \bahat      {\hat{\bf a}}
\newcommand \bB         {{\bf B}}

\newcommand \bE         {{\bf E}}

\newcommand \bH         {{\bf H}}
\newcommand \bJ         {{\bf J}}

\newcommand \bkhat      {\hat{\bf k}}

\newcommand \bznewhat      {\hat{\bf z}}
\newcommand \bxnewhat      {\hat{\bf x}}
\newcommand \bynewhat      {\hat{\bf y}}

\newcommand \beq        {\begin{equation}}
\newcommand \beqa	{\begin{eqnarray}}

\newcommand \calP       {{\cal P}}
\newcommand \cm         {\,{\rm cm}}

\newcommand \eeq	{\end{equation}}
\newcommand \eeqa	{\end{eqnarray}}

\newcommand \falign     {f_{\rm align}}
\newcommand \GHz        {\,{\rm GHz}}
\newcommand \gtsim	{\gtrsim}		 %apj version
\newcommand \Ha 	{{\rm H}}

\newcommand \K  	{\,{\rm K}}

\newcommand \lambdap    {\lambda_{\rm p}}

\newcommand \ltsim	{\lesssim}		 %apj version

\newcommand \MJy        {\,{\rm MJy}}

\newcommand \nH         {n_{\rm H}}
\newcommand \NH         {N_{\rm H}}
\newcommand \obs        {{\rm obs}}
\newcommand \MPFA        {{\rm MPFA}}
\newcommand \Phisbg     {\Phi_{\rm Ad}^{\rm MPFA}}
\newcommand \Piobs      {\Pi_{\rm obs}}
\newcommand \poro       {\calP}

\newcommand \Qabs       {Q_{\rm abs}}
\newcommand \Qpolabs    {Q_{\rm pol,abs}}
\newcommand \Qpolext    {Q_{\rm pol,ext}}

\newcommand \sr  	{\,{\rm sr}}

\newcommand \Tgr        {T_{\rm gr}}

\newcommand \xtimes     {{\!\,\times\!\,}}

\newcommand \achar      {a_{\rm char}}
\newcommand \aeff       {a_{\rm eff}}

\newcommand \Cabs       {C_{\rm abs}}
\newcommand \Cran       {C_{\rm ran}^{\rm MPFA}}
\newcommand \Cpolabs    {C_{\rm pol,abs}}
\newcommand \Cpolext    {C_{\rm pol,ext}}

\newcommand \pol        {{\rm pol}}

\newcommand{\btdnote}[1]{}
\newcommand{\btdomit}[1]{}

\newcommand{\oldtext}[1]{}

\countdef\decade=200
\advance\decade by \year
\countdef\hours=201
\advance\hours by \time
\divide\hours by 60
\countdef\mins=202
\advance\mins by \hours
\multiply\mins by 60
\multiply\hours by 100
\countdef\miltime=203
\advance\miltime by \hours
\advance\miltime by \time
\advance\miltime by -\mins

\begin{document}

\title{%
%------------- enable for labelling preprint ---------------------------
%        \vspace*{-3.0em}
%        {\normalsize\rm {\it The Astrophysical Journal}, {\bf 780}, 172 (2014 Jan.\ 10)}\\ 
%        \vspace*{1.0em}
%-----------------------------------------------------------------------
        {\bf Using the Starlight Polarization Efficiency Integral
        to Constrain Shapes and Porosities of Interstellar Grains}
%------------- disable for preprint
%	\\{\small draft: v\versnum\
%	  \today}%
        \\{\small\it submitted to ApJ}
%----------------------------------
	}

\author[0000-0002-0846-936X]{B. T. Draine}
\affiliation{Department of Astrophysical Sciences,
Princeton University, Princeton, NJ 08544-1001, USA}
\author[0000-0001-7449-4638]{Brandon S. Hensley}
\affiliation{Department of Astrophysical Sciences,
Princeton University, Princeton, NJ 08544-1001, USA}
\affiliation{Spitzer Fellow}
%\affiliation{Jet Propulsion Laboratory, California Institute of Technology, 4800
%Oak Grove Drive, Pasadena, CA 91109, USA}

\correspondingauthor{B. T. Draine}
\email{draine@astro.princeton.edu}

\begin{abstract}
  We present a new method for using
  the observed starlight polarization and polarized submm emission
  to constrain the shapes and porosities of 
  interstellar grains.
  We present the modified picket fence approximation (MPFA), and verify that
  it is sufficiently accurate for modeling
  starlight polarization.
  We introduce 
  the {\it starlight polarization integral\,} $\Piobs$ as a measure of
  overall strength of the observed polarization of starlight,
  and the {\it starlight polarization efficiency integral\,} $\Phi$
  to characterize the effectiveness of different grain types
  for producing polarization of starlight.
  The starlight polarization integral $\Piobs$ determines the
  mass-weighted alignment $\langle\falign\rangle$ of the grains.
  Approximating the aligned grains in the interstellar medium 
  as spheroids,
  we use $\Piobs/\Phi$ to show that the observed starlight polarization
  constrains the grains to have a minimum degree of asphericity.
  For porosity $\poro=0$, the minimum axial ratio is $\sim$1.4 for oblate
  spheroids, or $\sim$1.8 for prolate spheroids.
  If the grains are porous, more extreme axial ratios are required.
  The same grains that produce the starlight polarization are able to
  provide the observed polarized emission at submm wavelengths,
  but with further limits on shape and porosity.
  Porosities $\poro\gtsim0.75$ are ruled out.
  If interstellar grains can be approximated by astrodust spheroids, 
  we predict the ratio of 10$\micron$ polarization to starlight
  polarization $p_V$: $p(10\micron)/p_V=0.222\pm0.026$.
  For Cyg OB2-12 we predict $p(10\micron)=(2.1\pm0.3)\%$, which should be
  observable.

\end{abstract}

\let\svthefootnote\thefootnote
\let\thefootnote\relax\footnote{\textcopyright 2020.  All rights reserved.}
\let\thefootnote\svthefootnote

\section{Introduction
         \label{sec:intro}}

The polarization of starlight was discovered
serendipitously in 1948 \citep{Hiltner_1949b,Hall_1949} and
immediately attributed to linear dichroism of the interstellar medium 
arising from aligned dust grains.
Despite many decades of observational and theoretical study,
the physics of grain alignment remains uncertain
\citep[see the review by][]{Andersson+Lazarian+Vaillancourt_2015}.
In addition to polarizing starlight 
from the far-ultraviolet to the near-infrared,
the aligned grains also emit
polarized submm radiation and, on suitable sightlines,
produce measurable polarization of the 10$\micron$
silicate feature in absorption
\citep[e.g.][]{Dyck+Capps+Forrest+Gillett_1973,
               Wright+Aitken+Smith+etal_2002}.
Irrespective of the aligning mechanisms, 
observations of polarized extinction and emission can be used to constrain
possible grain shapes and degree of alignment.

For a given grain shape, size, 
and wavelength $\lambda$,
cross sections for absorption and scattering
depend on the grain orientation 
relative to the propagation direction $\bkhat$ and
polarization $\bE$ of the incident radiation.
These cross sections should be averaged over the actual 
distribution of grain
orientations.  Because this is numerically challenging, 
a simplified approach 
that is
often referred to as the ``picket fence approximation''
\citep[PFA;][]{Dyck+Beichman_1974} has sometimes been used.
%but the accuracy of
%the PFA
%at optical wavelengths does not
%appear to have been previously examined.
In the present paper we use
a slightly different approximation, which we
refer to as the modified PFA (MPFA).
The MPFA was employed in the infrared by \citet{Lee+Draine_1985},
and used to compute starlight polarization by 
\citet{Draine+Allaf-Akbari_2006} and
\citet{Draine+Fraisse_2009}.
Here we test the MPFA by comparing to actual 
averages for selected distributions of orientations.
We find that the MPFA is sufficiently accurate
for most purposes.

In the MPFA, the alignment of interstellar grains is characterized by the
size-dependent fractional alignment $\falign(a)$, where $a$ is a measure of
the grain size, and a corresponding mass-weighted fractional
alignment $\langle\falign\rangle$.
The polarization at long wavelengths $\lambda\gtsim 10\micron$ is
determined by the grain shape and $\langle\falign\rangle$.

Previous studies of the wavelength-dependent polarization of starlight have
solved for both the grain size distribution and $\falign(a)$
\citep{Kim+Martin_1995b,
       Draine+Allaf-Akbari_2006,
       Draine+Fraisse_2009,
       Guillet+Fanciullo+Verstraete+etal_2018}, 
but this approach can be 
very time-consuming.
Here we show that for determination of $\langle\falign\rangle$
this step can be bypassed.
The {\it starlight polarization efficiency integral\,} $\Phi$ -- an
integral over the (theoretically-calculated)
polarization cross section per grain volume --
can be combined with the observed
{\it starlight polarization integral\,} $\Piobs$ --
an integral over the observed starlight polarization -- to determine
$\langle\falign\rangle$ 
without solving for the size distribution or $\falign(a)$.
An upper bound on $\langle\falign\rangle$ then leads to a lower bound on the
axial ratio of the grains.
The polarized submm emission observed by {\it Planck} provides
an addditional constraint, further limiting allowed values of
{\it both} porosity
{\it and} grain shape.

The paper is organized as follows.
In Section \ref{sec:orientational averaging} we review
averaging over spinning, precessing
spheroids, and in Section \ref{sec:MPFA} we present the MPFA
for estimating orientational averages.
The MPFA is tested in Section \ref{sec:MPFA test}.
In Sections \ref{sec:Phi} and \ref{sec:polint} we review the observed
polarization of starlight and define the 
starlight polarization efficiency integral
$\Phi$ for dust grains of specified shape and composition.
In Section \ref{sec:lambda_peff} we define the effective polarizing wavelength
$\lambda_{\rm p,eff}$, and we deduce the
characteristic size $a_{\rm char}$ of the grains responsible for
the observed polarization of starlight.
In Section \ref{sec:falign}
we show how the mass-weighted alignment fraction
$\langle f_{\rm align}\rangle$ can be estimated from the ratio
$\Piobs/\Phi$.

Approximating the grains by spheroids,
polarized submm emission from the aligned dust grains is discussed
in Section \ref{sec:polarized emission}; 
the strength of the observed polarized submm emission leads to a limited
domain of allowed shapes and porosities.
Polarization in the $10\micron$ silicate feature is discussed in
Section \ref{sec:10um pol}, where we calculate the 10$\micron$ polarization
per starlight polarization, 
and predict $p(10\micron)=(2.2\pm0.3)$\% toward Cyg OB2-12.
We discuss our conclusions in Section \ref{sec:discussion}.
Our principal results are summarized in Section \ref{sec:summary}.

% sec 2
\section{\label{sec:orientational averaging}
         Orientational Averaging for Spinning, Precessing Spheroids}

Consider grains 
with rotational symmetry around a symmetry axis $\bahat$,
and reflection symmetry through a plane perpendicular to $\bahat$
(e.g., spheroids or cylinders).
For simplicity, we will discuss spheroids.

Let photons be propagating in direction $\bkhat=\bznewhat$, 
and
let $\Theta$ be the angle between $\bkhat$ and $\bahat$.
Cross sections for absorption or scattering of light with
polarization $\bE$
depend on $\Theta$ and the orientation of
$\bE$ with respect to $\bahat$.
Let $C_E(\Theta)$ be the cross section for $\bE$ in the $\bkhat-\bahat$
plane, and $C_H(\Theta)$ be the cross section for the magnetic field $\bH$  in
the $\bkhat-\bahat$ plane.
The cross sections for
incident light polarized in the $\bxnewhat$ and
$\bynewhat$ directions are
\beqa
C_x(\Theta) &=& 
\frac{(\bahat\cdot\bxnewhat)^2}{(\bahat\cdot\bxnewhat)^2 + (\bahat\cdot\bynewhat)^2}
C_E(\Theta)
+
\frac{(\bahat\cdot\bynewhat)^2}{(\bahat\cdot\bxnewhat)^2 + (\bahat\cdot\bynewhat)^2}
C_H(\Theta)
\\
C_y(\Theta) &=& 
\frac{(\bahat\cdot\bynewhat)^2}{(\bahat\cdot\bxnewhat)^2 + (\bahat\cdot\bynewhat)^2}
C_E(\Theta)
+
\frac{(\bahat\cdot\bxnewhat)^2}{(\bahat\cdot\bxnewhat)^2 + (\bahat\cdot\bynewhat)^2}
C_H(\Theta)
~~~,
\eeqa
where $C$ can denote cross section for absorption, scattering, or extinction.

A grain with angular momentum $\bJ$ undergoes
tumbling motion, with its angular velocity $\bomega$ nutating
around $\bJ$.  If the grain is treated as a rigid body,
the torque-free motion of the
principal axes in space is given by the Euler equations
\citep{Landau+Lifshitz_1976}.
If the grain is non-rigid, with internal processes
(e.g., viscoelasticity) that can exchange rotational kinetic
energy $E_{\rm rot}$
with the vibrational degrees of freedom (heat content of the grain),
the motion becomes more complex
\citep{Lazarian+Roberge_1997,
Lazarian+Draine_1997,
Lazarian+Draine_1999a,
Weingartner+Draine_2003,
Weingartner_2009,
Kolasi+Weingartner_2017}.

However, if grains are rotating suprathermally, with $E_{\rm rot}\gg kT$, then
it is expected that dissipational mechanisms within the grain will
act to minimize $E_{\rm rot}$ at fixed $\bJ$, causing the grain
to spin around the principal axis of largest moment of inertia
\citep{Purcell_1975}.
An oblate spheroid will then have its symmetry axis aligned with $\bJ$;
a prolate spheroid will spin with its symmetry axis perpendicular
to $\bJ$.
For simplicity of presentation, we will assume this limit, where the
orientational averages for a grain
depend only on the distribution of the directions of $\bJ$.

If the grain material has unpaired electrons, the Barnett effect
(partial alignment of the electron spins with the angular velocity $\bomega$)
contributes a magnetic moment antiparallel to $\bomega$
\citep{Dolginov+Mytrophanov_1976,Purcell_1979}.
If the grain has positive (negative) net charge, 
the Rowland effect contributes
a magnetic moment parallel (antiparallel) 
to $\bomega$ \citep{Dolginov_1968,Martin_1971}.
The net magnetic moment is typically
large enough that the grain angular momentum
$\bJ$ precesses
relatively rapidly
around the local magnetic field $\bB_0$.
The problem of orientational averaging then reduces to specifying
the direction of $\bB_0$ and the distribution of the
``cone angle'' $\psi$ between $\bJ$ and $\bB_0$,
and, for each $\psi$, averaging over grain rotation around $\bJ$
and precession of $\bJ$ around $\bB_0$.

Let $\bB_0$ lie in the $\bznewhat-\bxnewhat$ plane, with
$\gamma$ the angle between $\bB_0$ and the direction of propagation $\bznewhat$.
Let $\zeta$ be the precession angle, and
let angle $\beta$ measure rotation of the grain around $\bomegahat$.
Let $p_{\rm align}(\psi)d\psi$ 
be the fraction of the grains having alignment angle
$\psi$ in $[\psi,\psi+d\psi]$.
The limit of perfect alignment is $p_{\rm align}(\psi)\rightarrow \delta(\psi)$,
where $\delta$ is the usual delta function.
The limit of random orientation is $p_{\rm align}(\psi)\propto \sin\psi$.

\subsection{Oblate Spheroids}

For oblate spheroids spinning with $\bahat\parallel\bomega$
we can ignore the rotation angle $\beta$:
\beqa \label{eq:obl1}
\bahat\cdot\bxnewhat &=& \sin\gamma\cos\psi - \cos\gamma\sin\psi\cos\zeta
\\ \label{eq:obl2}
\bahat\cdot\bynewhat &=& \sin\psi\sin\zeta
\\ \label{eq:obl3}
\cos\Theta \equiv \bahat\cdot\bznewhat 
&=& \cos\gamma\cos\psi + \sin\gamma\sin\psi\cos\zeta
~,
\eeqa
and orientationally-averaged cross sections are given by
\beq \label{eq:<C> oblate}
\left\langle C \right\rangle = 
\int_0^{\pi/2} p_{\rm align}(\psi)d\psi \int_0^{2\pi} 
\frac{d\zeta}{2\pi} C(\zeta,\psi)
~.
\eeq
\subsection{Prolate Spheroids}

For prolate grains spinning with $\bahat\perp\bomega$:
\beqa \label{eq:pro1}
\bahat\cdot\bxnewhat &=&
               -\cos\gamma\sin\zeta\sin\beta -
                (\cos\gamma\cos\psi\cos\zeta + \sin\gamma\sin\psi)\cos\beta
\\ \label{eq:pro2}
\bahat\cdot\bynewhat &=& -\cos\zeta\sin\beta + \cos\psi\sin\zeta\cos\beta
\\ \label{eq:pro3}
\cos\Theta \equiv \bahat\cdot\bznewhat 
&=& \sin\gamma\sin\zeta\sin\beta +
               (\sin\gamma\cos\psi\cos\zeta - \cos\gamma\sin\psi)\cos\beta
~,
\eeqa
and
\beq \label{eq:<C> prolate}
\left\langle C \right\rangle =
\int_0^{\pi/2}p_{\rm align}(\psi)d\psi
\int_0^{2\pi} \frac{d\zeta}{2\pi}
\int_0^{2\pi} \frac{d\beta}{2\pi}
C(\beta,\zeta,\psi)
~.
\eeq

% sec 3
\section{\label{sec:MPFA}
            Modified Picket Fence Approximation (MPFA)}

Equations (\ref{eq:<C> oblate}) and (\ref{eq:<C> prolate}) are exact
for spheroids (or any other shape with axial symmetry and
reflection symmetry through the equatorial plane), assuming the spin
axis to be parallel to the principal axis of largest moment of inertia.
However, evaluation of the integrals (\ref{eq:<C> oblate}) and
(\ref{eq:<C> prolate}) requires 
$C_E(\Theta)$ and $C_H(\Theta)$ for many values of $\Theta$, which
can be computationally demanding.
 
\citet{Dyck+Beichman_1974} introduced what they called the ``picket fence
approximation,'' which consisted of assuming that some fraction of
the grains were perfectly aligned to produce polarization in a preferred
direction, while the remainder of the particles were randomly oriented.
Here we will adopt a somewhat different approximation, which we will refer to
as the ``modified picket fence approximation'' (MPFA).
The MPFA is equally simple, and leads to similar equations, 
but allows us to explicitly discuss polarization by a population of
partially-aligned, spinning and precessing grains.
While the MPFA has been used previously
\citep{Lee+Draine_1985,Draine+Allaf-Akbari_2006,Draine+Fraisse_2009,
Vandenbroucke+Baes+Camps_2020}, it is helpful to review it here.

For simplicity, we limit the discussion to grains with axial 
symmetry and reflection symmetry.
We will discuss spheroids as a specific example; the results are easily
extended to other cases, e.g., cylinders.
The MPFA then consists of
assuming that the
cross sections $C_E$ and $C_H$ depend 
linearly on $\cos^2\Theta$:
\beqa \label{eq:ed 1}
C_E(\Theta) &\approx& \cos^2\!\Theta\, C_E(0) + \sin^2\!\Theta\, C_E(90^\circ)
\\ \label{eq:ed 2}
C_H(\Theta) &\approx& \cos^2\!\Theta\, C_H(0) + \sin^2\!\Theta\, C_H(90^\circ)
~~~,
\eeqa
which further simplifies because $C_E(0)=C_H(0)$.

For a grain of volume $V$, we characterize the size by
the radius of an equal volume sphere,
\beq
\aeff\equiv\left(\frac{3V}{4\pi}\right)^{1/3}
~~.
\eeq
In the electric dipole limit ($\aeff/\lambda \rightarrow 0$),
the linear dependence of $C_E$ and $C_H$ on $\cos^2\Theta$ in
(\ref{eq:ed 1},\ref{eq:ed 2}) is exact\footnote{%
   In the electric dipole limit, interaction of a grain with an applied
   $\bE$ field is characterized by the complex polarizability tensor 
   $\alpha_{ij}$ \citep{Draine+Lee_1984}.
   Choosing coordinates where $\alpha_{ij}$ is diagonalized,
   $$
   \Cabs=\frac{4\pi\omega}{c}
   \frac{\sum_{j=1}^3|E_j|^2{\rm Im}(\alpha_{jj})}{\sum_{j=1}^3 |E_j|^2}
   ~,
   $$
   and Eqs.\ (\ref{eq:ed 1},\ref{eq:ed 2}) follow directly.
   }
and the MPFA is, therefore, exact in that limit.

We define the alignment efficiency
\beq \label{eq:falign}
\falign(\aeff)\equiv \frac{3}{2}
\left(\langle \cos^2\psi\rangle - \frac{1}{3}\right)
~,
\eeq
and MPFA polarization cross sections
\beqa \nonumber 
{\rm oblate:}~\Cpolext^\MPFA(\lambda) &\equiv& 
\frac{1}{2}\left[C_{{\rm ext},H}(90^\circ) - C_{{\rm ext},E}(90^\circ)\right]
\\ \label{eq:CpolMPFA oblate}
&\equiv& \frac{1}{2}\left[C_{\rm ext}(\bkhat\perp\bahat,\bE\perp\bahat)-
                         C_{\rm ext}(\bkhat\perp\bahat,\bE\parallel\bahat)\right]
~~~~
\\ \nonumber
{\rm prolate:}~ \Cpolext^\MPFA(\lambda) 
&\equiv& 
\frac{1}{4}\left[C_{{\rm ext},E}(90^\circ) - 
                 C_{{\rm ext},H}(90^\circ)\right]
\\ \label{eq:CpolMPFA prolate}
&\equiv& \frac{1}{4}\left[ C_{\rm ext}(\bkhat\perp\bahat,\bE\parallel\bahat)-
                         C_{\rm ext}(\bkhat\perp\bahat,\bE\perp\bahat)\right]
.~~~
\eeqa
Orientational averages are discussed in Appendix \ref{app:pfa}. 
From Eq.\ (\ref{eq:<Cx-Cy>MPFA}), we have
\beq \label{eq:MPFA}
\frac{1}{2}\langle C_{{\rm ext},y} -C_{{\rm ext},x} \rangle^\MPFA 
= \Cpolext^\MPFA(\lambda) \falign(\aeff) \sin^2\gamma
~~~.
\eeq
The polarization-averaged extinction cross section is also dependent
on $\gamma$ and $\falign$:
\beqa
{\rm oblate:}~
\frac{\langle C_{{\rm ext},x}\!+\!C_{{\rm ext},y} \rangle^\MPFA}{2}
&=& \Cran + \falign\left(\cos^2\gamma-\frac{1}{3}\right)
\!
\left[C_{{\rm ext},E}(0^\circ)-
\frac{C_{{\rm ext},E}(90^\circ)\!+\!C_{{\rm ext},H}(90^\circ)}{2}\right]
\\
{\rm prolate:}~
\frac{\langle C_{{\rm ext},x}\!+\!C_{{\rm ext},y} \rangle^\MPFA}{2}
&=& \Cran - \frac{\falign}{2}\left(\cos^2\gamma-\frac{1}{3}\right)
\!
\left[C_{{\rm ext},E}(0^\circ)-
\frac{C_{{\rm ext},E}(90^\circ)\!+\!C_{{\rm ext},H}(90^\circ)}{2}\right]\!,
~~~~~~
\eeqa
where
\beq
\Cran \equiv 
\frac{1}{3}\left[C_{{\rm ext},E}(0)+C_{{\rm ext},E}(90^\circ)+
C_{{\rm ext},H}(90^\circ)\right]
~.~~~~
\eeq
For finite values of $\aeff/\lambda$,
the cross sections $C_E(\Theta)$ and $C_H(\Theta)$
will have a more complex dependence on $\Theta$ than the simple
linear dependence on $\cos^2\Theta$
in Eqs.\ (\ref{eq:ed 1}) and (\ref{eq:ed 2}), but we anticipate that
orientational averaging may make $\langle C_x\rangle$ and
$\langle C_y \rangle$ relatively insensitive to deviations from the
assumed linear dependence on $\cos^2\Theta$ 
in Eqs.\ (\ref{eq:ed 1}) and (\ref{eq:ed 2}).
This conjecture is tested below.

% sec 4
\section{\label{sec:MPFA test}
         Testing the MPFA}

At wavelengths $\lambda > 8\micron$, submicron grains have
$2\pi\aeff/\lambda \ll 1$, 
%the electric dipole approximation is highly accurate, 
and the MPFA (Eqs.\ \ref{eq:ed 1}, \ref{eq:ed 2}) is known to be accurate,
as verified by \citet{Vandenbroucke+Baes+Camps_2020}.
However, 
at ``optical'' wavelengths where starlight polarization is measured, 
$2\pi\aeff/\lambda \gtsim 1$, and
the accuracy of the MPFA must be checked by comparison to
the exact averages in Eq.\ (\ref{eq:<C> oblate}) and
(\ref{eq:<C> prolate}).
 
When
$2\pi\aeff/\lambda \gtsim 0.5$,
calculation of scattering and absorption by shapes other than
spheres becomes challenging.
Electromagnetic
absorption and scattering by spheroids
can be treated using various approaches, including separation of variables
\citep{Asano+Yamamoto_1975,Voshchinnikov+Farafonov_1993},
the ``extended boundary condition method'' 
\citep{Waterman_1965,Mishchenko_2000,Mishchenko_2020}
(often referred to as the ``T-matrix method''),
the 
``generalized point matching method''
\citep{Al-Rizzo+Tranquilla_1995,Weingartner+Jordan_2008},
or the discrete dipole approximation \citep{Draine+Flatau_1994}.
Because the calculations are time-consuming (and become
numerically delicate when $2\pi\aeff/\lambda \gtsim 10$)
fast methods are desirable, but accuracy must be verified.

%%%%%%%%%%%%%%%%%%%%%%%%%%%%%%%%% f1 %%%%%%%%%%%%%%%%%%%%%%%%%%%%%%%%%
\begin{figure}[b]
\begin{center}
\includegraphics[angle=0,width=8.0cm,
                 clip=true,trim=0.5cm 5.0cm 0.5cm 0.5cm]%l b r t
{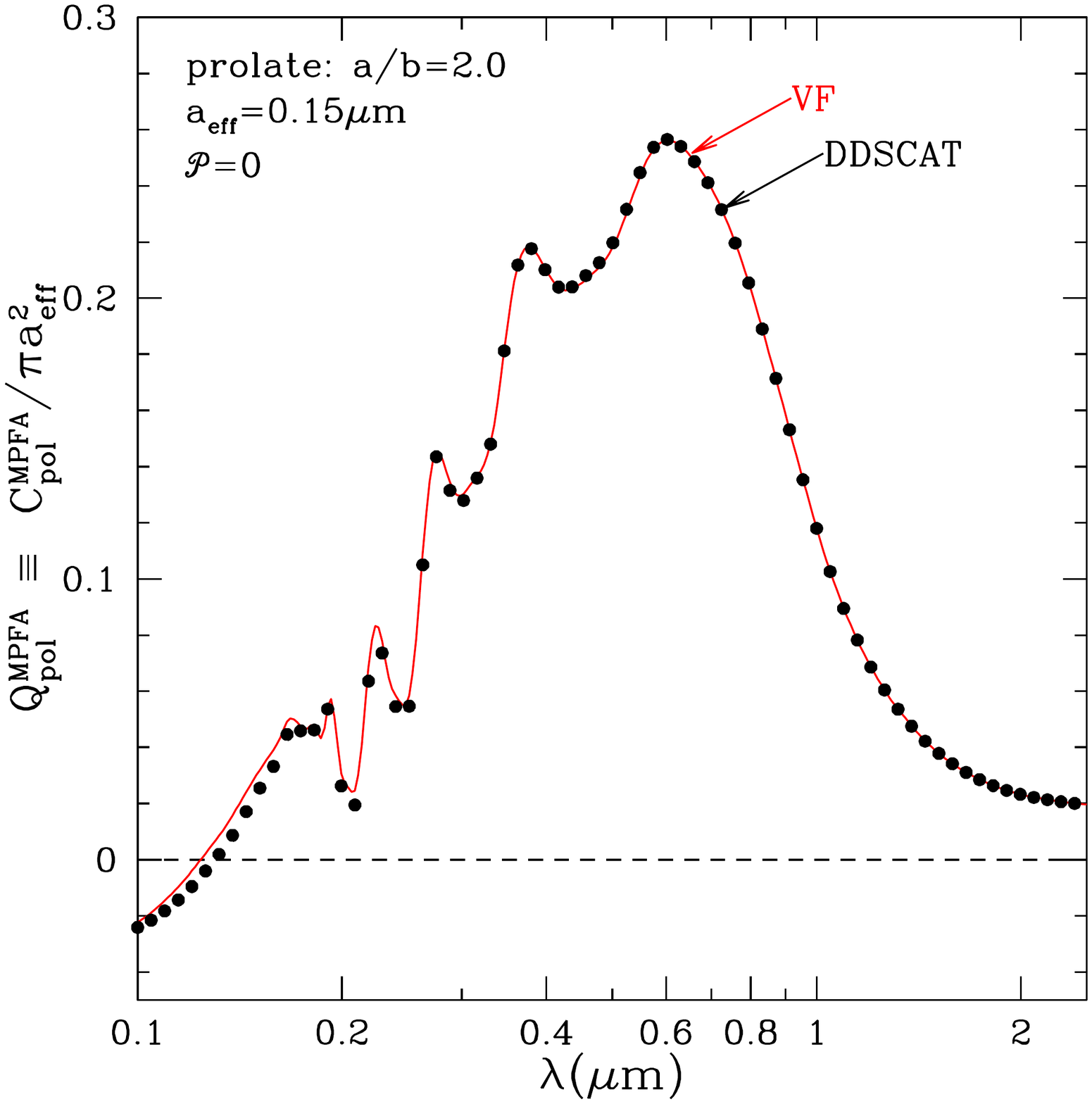}
\includegraphics[angle=0,width=8.0cm,
                 clip=true,trim=0.5cm 5.0cm 0.5cm 0.5cm]%l b r t
{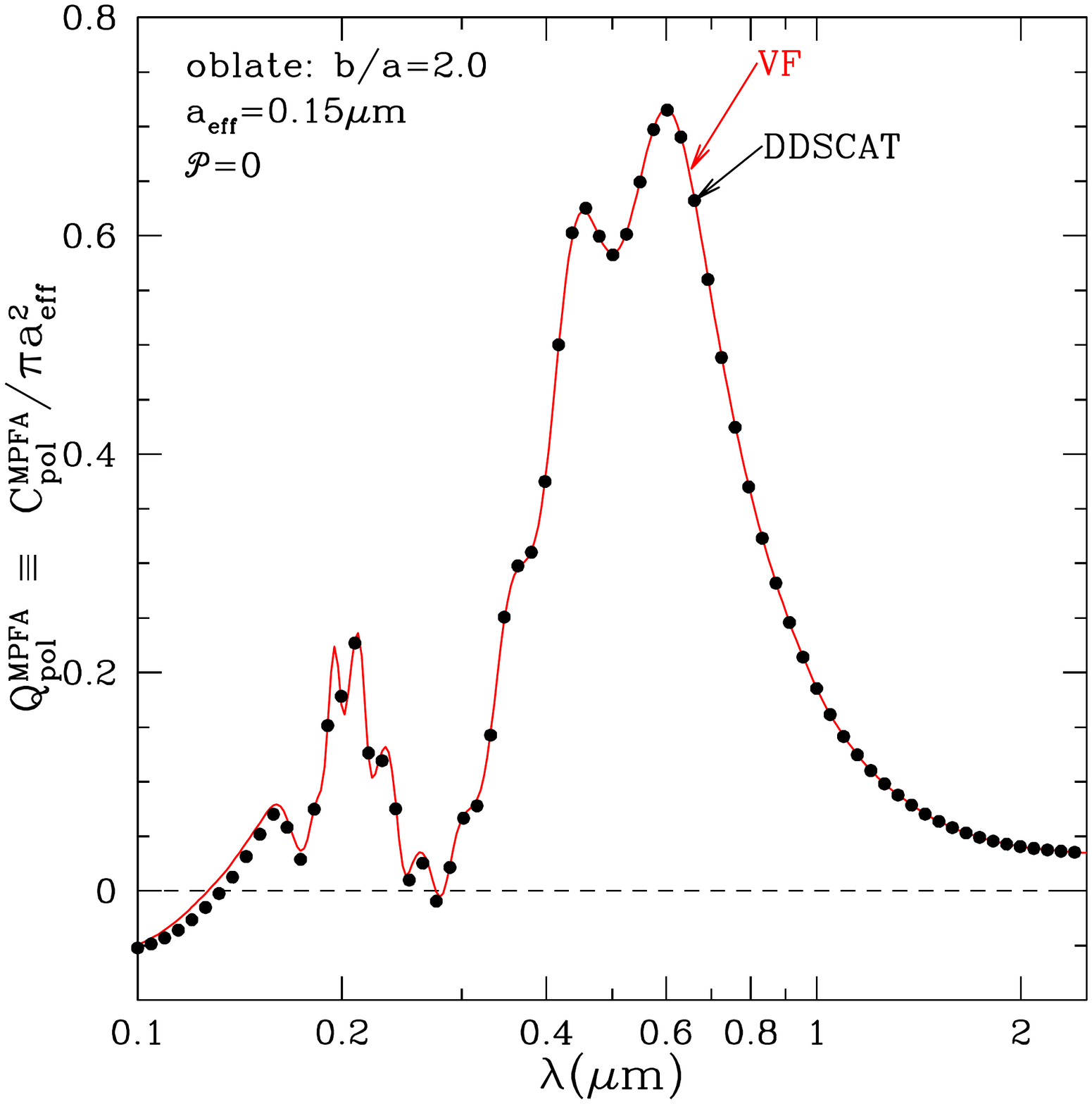}
\caption{\label{fig:VF vs DDA} \footnotesize
         Normalized polarization cross section
         for $\aeff=0.15\micron$ grains composed of silicate-bearing
         material with porosity $\poro=0$.
         Left: 2:1 prolate spheroid.
         Right: 2:1 oblate spheroid.
         Red curve: method of \citet{Voshchinnikov+Farafonov_1993} (VF).
         Black points: DDA calculations (DDSCAT).
         Agreement is excellent for both oblate and prolate cases.
         \btdnote{f1a.pdf, f1b.pdf}
         }
\end{center}
\vspace*{-0.3cm}
\end{figure}
%%%%%%%%%%%%%%%%%%%%%%%%%%%%%%% end f1 %%%%%%%%%%%%%%%%%%%%%%%%%%%%%%%%

We consider grains composed of ``astrodust'' material, using dielectric
functions derived by
\citet{Draine+Hensley_2021a}.\footnote{%
    These dielectric functions are available at \url{http://www.astro.princeton.edu/~draine/DH21Ad.html}}
Astrodust is assumed to consist of amorphous
silicate material (with nominal composition 
Mg$_{1.3}$Fe$_{0.3}$SiO$_{3.6}$) 
mixed with carbonaceous material and other
nonsilicate materials such as Al$_2$O$_3$, FeS, Fe$_3$O$_4$, and
possibly metallic Fe.
The silicate material itself accounts for
approximately 50\% of the mass of the astrodust material
(see \citet{Draine+Hensley_2021a} for discussion of likely constituents).
Astrodust grains may also contain voids, with volume filling factor
(i.e., porosity) $\poro$.
We consider porosities in the range $0\leq\poro\leq0.9$, and
various axial ratios $b/a$.
For each choice of $\poro$ and $b/a$, we use the appropriate
dielectric function obtained by \cite{Draine+Hensley_2021a}, based
on fits to the observed extinction toward Cyg OB2-12 
\citep{Hensley+Draine_2020} and estimates of the dust 
opacity based on observations at
other wavelengths \citep{Hensley+Draine_2021a}.

Absorption, scattering, and extinction cross sections
$C_E(0)=C(\bkhat\parallel\bahat)$,
$C_E(90^\circ)=C(\bkhat\perp\bahat,\bE\parallel\bahat)$,
and
$C_H(90^\circ)=C(\bkhat\perp\bahat,\bE\perp\bahat)$
have been calculated for spheroids
using the separation of variables method code
of \citet{Voshchinnikov+Farafonov_1993}.
Tabulated results for a range of grain sizes and shapes,
wavelengths $\lambda>912\Angstrom$,
and a range of porosities
are available at
\url{http://www.astro.princeton.edu/~draine/DH21Ad.html}.

Figure \ref{fig:VF vs DDA} shows the dimensionless polarization efficiency
\beq
\Qpolext^\MPFA \equiv \frac{\Cpolext^\MPFA}{\pi\aeff^2}
\eeq
for zero porosity ($\calP=0$)
astrodust grains with size $\aeff=0.15\micron$, as a function of vacuum
wavelength $\lambda$,
calculated using the separation of variables method
code written by \citet{Voshchinnikov+Farafonov_1993}
(VF)\footnote{%
   hom6\_5q, available from
   \url{http://www.astro.spbu.ru/DOP/6-SOFT/SPHEROID/1-SPH\_new/}
   }    
and
with the discrete dipole approximation (DDA) \citep{Draine+Flatau_1994}
using the public-domain code DDSCAT.\footnote{%
   DDSCAT 7.3.3, available from \url{http://www.ddscat.org}}

The DDSCAT calculations were done with $N=265848$ and $N=277888$
dipoles for $b/a=2$, and $b/a=0.5$, respectively.
Figure \ref{fig:VF vs DDA} shows that 
the DDA and the VF results
are in excellent agreement,
confirming that both methods are accurate.
We will use DDA calculations to test the accuracy of the MPFA for
several distributions of orientations.
For all cases, we assume that $\bomega$ precesses around $\bB_0$ and
we average over this precession;
we use
Eqs.\ (\ref{eq:obl1}--\ref{eq:<C> oblate}) for oblate grains, or
Eqs.\ (\ref{eq:pro1}--\ref{eq:<C> prolate}) for prolate grains.

\begin{figure}[t]
\begin{center}
\includegraphics[angle=0,width=8.0cm,
                 clip=true,trim=0.5cm 0.5cm 0.5cm 0.5cm]%l b r t
{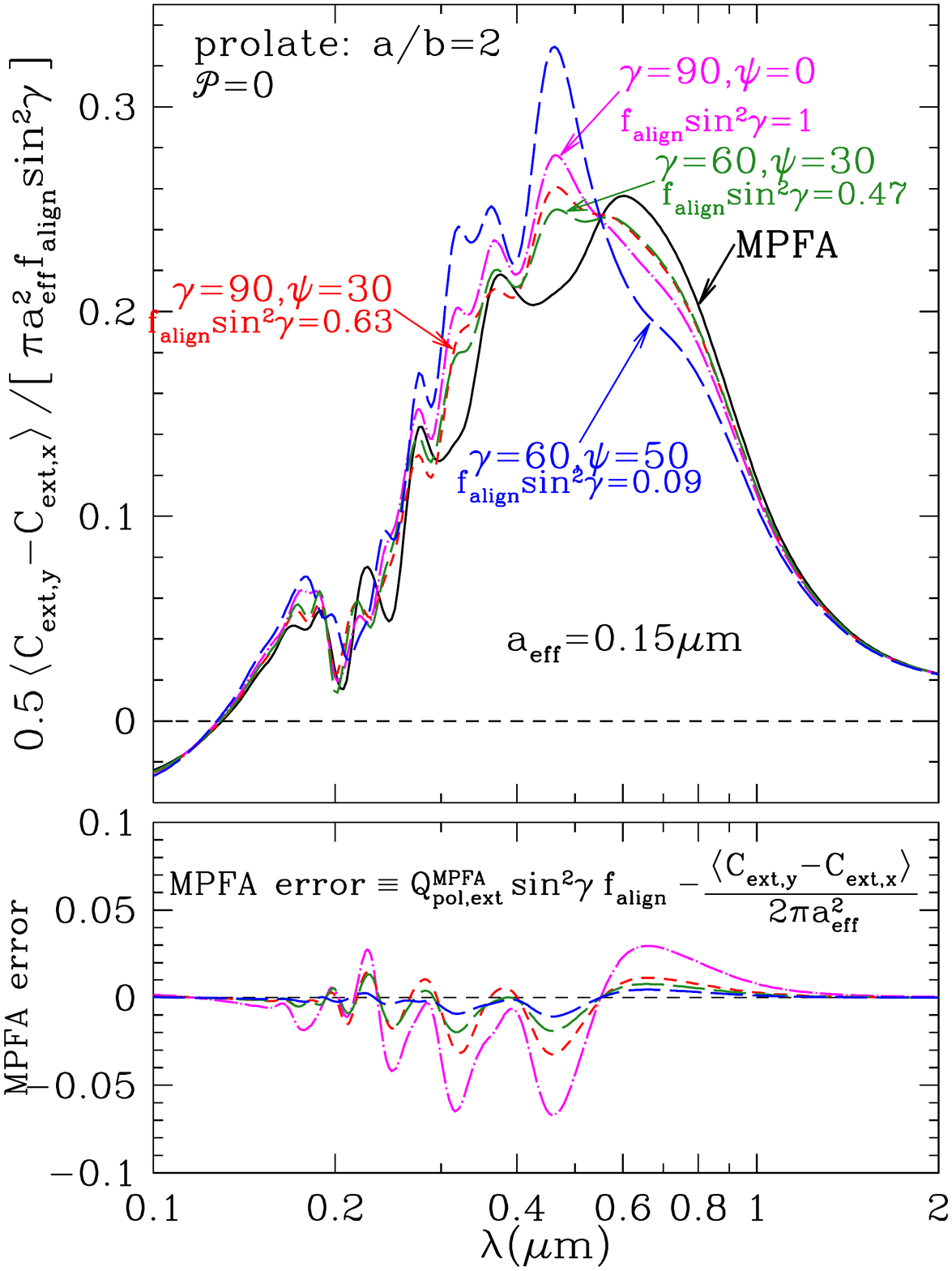}
\includegraphics[angle=0,width=8.0cm,
                 clip=true,trim=0.5cm 0.5cm 0.5cm 0.5cm]%l b r t
{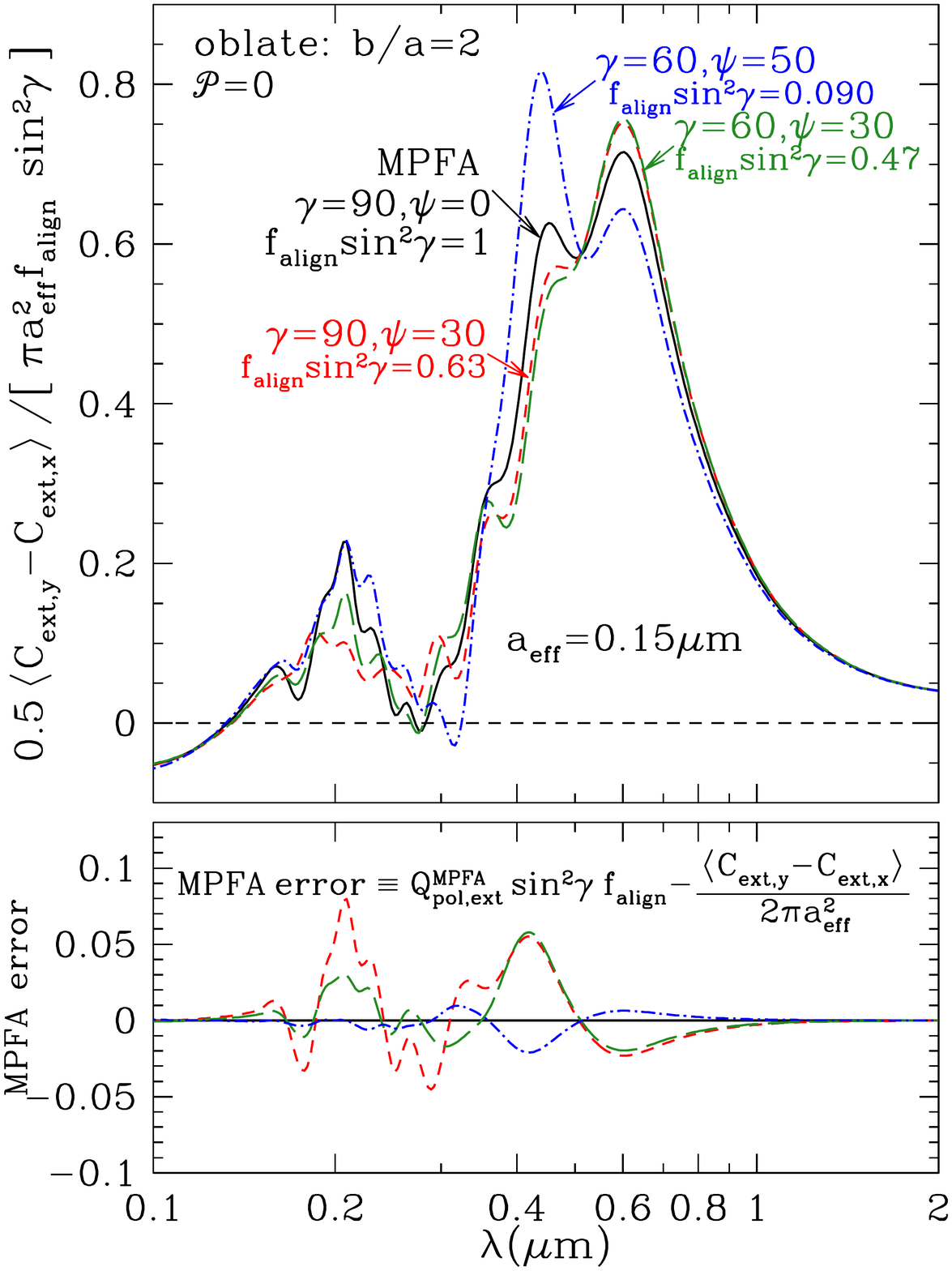}
\caption{\label{fig:orient avg} \footnotesize
         Polarization cross section $C_\pol\equiv(1/2)(C_x-C_y)$
         divided by $\pi \aeff^2 \falign\sin^2\gamma$,
         as a function of wavelength $\lambda$.
         Left: 2:1 prolate grains.
         Right: 2:1 oblate grains.
         Grains are spinning with $\bomega$ parallel
         to principal axis of largest moment of inertia, with
         $\bomega$ precessing around $\bB_0$.
         Results are shown
         for 4 different
         distributions of grain orientation, labelled by
         $\gamma=$ angle between $\bB_0$ and the l.o.s., and
         $\psi$ = angle between $\bomega$ and $\bB_0$.
         Curves labelled ``MPFA'' correspond to the
         ``modified picket fence'' approximation.
         Upper panels show the polarization efficiencies divided by
         $\falign\sin^2\gamma$ for different distributions.
         Lower panels show the error in 
         $C_\pol/\pi\aeff^2\falign\sin^2\gamma$ if the MPFA is used.
         Note that the errors are both positive and negative, and will tend
         to average out for size distributions.
         \btdnote{f2a.pdf, f2b.pdf}
         }
\end{center}
\vspace*{-0.3cm}
\end{figure}

Figure \ref{fig:orient avg} shows orientationally-averaged
optical polarization
cross sections as a function of wavelength $\lambda$
for four different orientation distributions:
\begin{enumerate}
\item $\bB_0\perp\bkhat$ ($\gamma=90^\circ$)
and perfect (spinning) alignment ($\psi=0^\circ$).
\item $\bB_0\perp\bkhat$ ($\gamma=90^\circ$) 
and $\psi=30^\circ$ between $\bomega$ and $\bB_0$.
\item $\gamma=60^\circ$ between $\bB_0$ and $\bkhat$, and
$\psi=30^\circ$ between $\bomega$ and $\bB_0$.
\item $\gamma=60^\circ$ between $\bB_0$ and $\bkhat$, and
$\psi=50^\circ$ between $\bomega$ and $\bB_0$.
\end{enumerate}
To evaluate the
orientational averages $\langle C_{{\rm ext},x}\rangle$ and 
$\langle C_{{\rm ext},y}\rangle$
using either Eq.\ (\ref{eq:<C> oblate}) or (\ref{eq:<C> prolate}),
we first calculate $C_{{\rm ext},E}(\Theta)$ and 
$C_{{\rm ext},H}(\Theta)$ for 11 values of
$\Theta$ between $0$ and $\pi/2$.
We then interpolate to obtain 
$C_{{\rm ext},x}$ and
$C_{{\rm ext},y}$ at values of $\Theta$ needed for the integral.

%%%%%%%%%%%%%%%%%%%%%%%%%%%%%%%%% f2 %%%%%%%%%%%%%%%%%%%%%%%%%%%%%%%
%%%%%%%%%%%%%%%%%%%%%%%%%%%%%%% end f2 %%%%%%%%%%%%%%%%%%%%%%%%%%%%%%%%%%
For each orientation distribution, the orientationally-averaged polarization
cross sections 
$\frac{1}{2}\langle C_{{\rm ext},y}-C_{{\rm ext},x}\rangle$
are normalized by dividing by
$\pi\aeff^2\falign\sin^2\!\gamma$.
Figure \ref{fig:orient avg} shows that these four distributions have
very similar normalized optical polarization profiles, for both the oblate and
prolate shapes.
The polarization cross sections calculated with the MPFA are also shown. 
% For oblate spheroids, the case
% $\gamma=90^\circ,\psi=0^\circ$ 
% has $\langle \Cpolext\rangle = \Cpolext^\MPFA$); for other cases,
% $\langle \Cpolext\rangle$ is close to the MPFA, although differing in detail.

The MPFA is exact in the limit $\lambda\rightarrow\infty$.
For $\aeff=0.15\micron$ the MPFA is quite accurate for $\lambda > 1\micron$,
($2\pi\aeff/\lambda < 1$), as shown in Fig.\ \ref{fig:orient avg}.

For oblate spheroids
the MPFA gives the exact result
for the case $\gamma=90^\circ$ and
$\psi=0^\circ$ (perfect alignment), 
but for other cases the orientational averages depart
from the MPFA for $\lambda \ltsim 1\micron$ 
($2\pi\aeff/\lambda \gtsim 1$).
The modest oscillatory behavior seen in Fig.\ \ref{fig:orient avg}
will be suppressed when averaging over realistic distributions of
$\aeff$ and $\psi$,
and we expect the MPFA to provide a good aproximation after such averaging.
We confirm this below.

% sec 5
\section{\label{sec:Phi}
         Starlight Polarization Efficiency Integral $\Phi$}

For each assumed orientation distribution $p_{\rm align}(\psi)$
and magnetic field orientation $\gamma$ we define
a dimensionless {\it starlight polarization efficiency integral}
\beq \label{eq:Phi def}
\Phi(\aeff,b/a,p_{\rm align},\gamma) \equiv 
\frac{1}{\falign(\aeff)\sin^2\!\gamma}
\int_{\lambda_1}^{\lambda_2} 
\frac{1}{2}\frac{\langle C_{{\rm ext},y}-C_{{\rm ext},x}\rangle}{V} d\lambda
~~~,
\eeq
where $\falign$ is given by Eq.\ (\ref{eq:falign}),
$V\equiv (4\pi/3)\aeff^3$ is the grain volume,
and $\langle C_{{\rm ext},y}-C_{{\rm ext},x}\rangle$ 
is the orientational average
calculated from Eq.\ (\ref{eq:<C> oblate}) or (\ref{eq:<C> prolate}).
We set
$\lambda_1=0.15\micron$ and $\lambda_2=2.5\micron$.
The methodology used here is insensitive to the exact choice of $\lambda_1$
and $\lambda_2$, provided only that they include the wavelength range
where starlight polarization is observed to be strong.

%%%%%%%%%%%%%%%%%%%%%%%%%%%%%%%% f3 %%%%%%%%%%%%%%%%%%%%%%%%%%%%%%%%%%%%%
\begin{figure}[ht]
\begin{center}
\includegraphics[angle=0,width=8.0cm,
                 clip=true,trim=0.5cm 0.5cm 0.5cm 0.5cm]%l b r t
{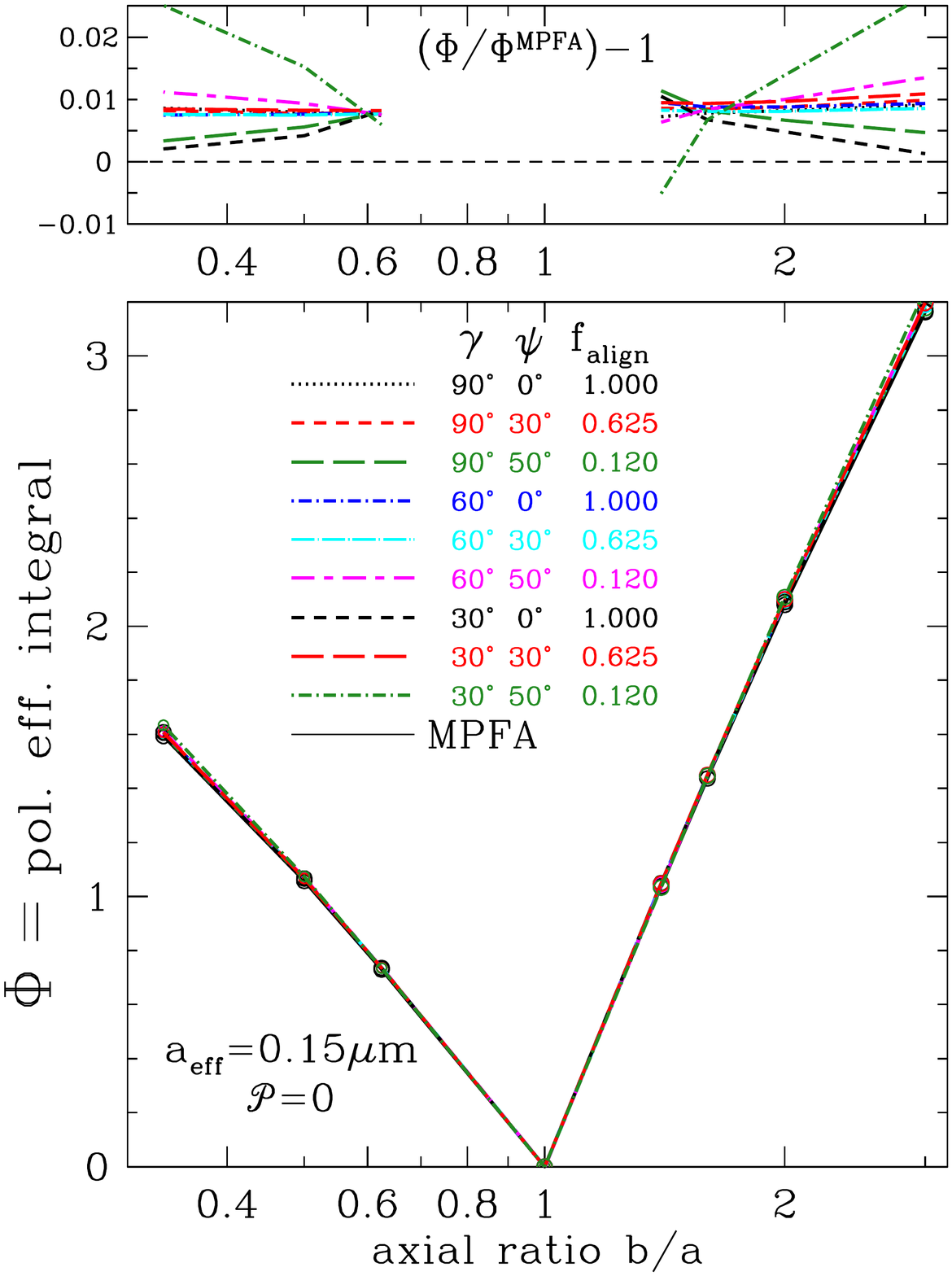}
\includegraphics[angle=0,width=8.0cm,
                 clip=true,trim=0.5cm 0.5cm 0.5cm 0.5cm]%l b r t
{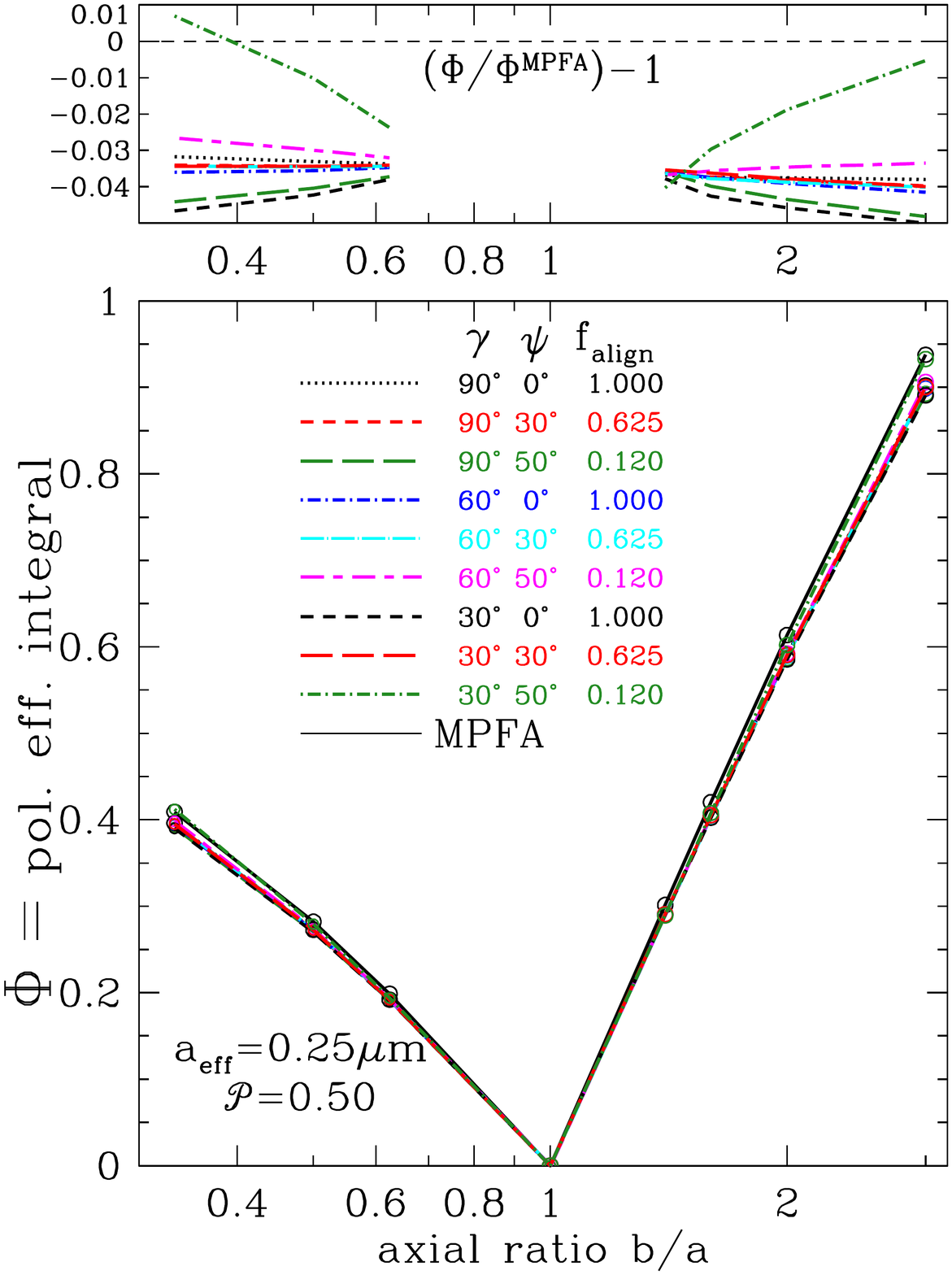}
\caption{\footnotesize \label{fig:Phi}
     Lower panels: Starlight polarization efficiency integral ${\Phi}$
     (Eq.\ \ref{eq:Phi def})
     for spheroids 
     as a function of axial ratio $b/a$, for various
     orientations (see text)
     for $\lambda_1=0.15\micron$, $\lambda_2=2.5\micron$.
     $\gamma$ is the angle between $\bB_0$ and the line of sight;
     $\psi$ is the angle between $\bomega$
     and $\bB_0$, with $\bomega$ precessing around $\bB_0$.
     Grains are ``astrodust'' (see text) with porosity $\calP=0$ (left)
     and $\calP=0.5$ (right).
     $\Phi$ is nearly independent
     of both $\gamma$ and the degree of alignment, 
     thus the MPFA is an
     excellent approximation.
     Upper panels: fractional error $(\Phi/\Phi^\MPFA)-1$ for different
     values of $\gamma$ and alignment function $p_{\rm align}$.
     The MPFA is generally accurate to within a few percent.
     \btdnote{f3a.pdf,f3b.pdf}
     }
\vspace*{-0.3cm}
\end{center}
\end{figure}
%%%%%%%%%%%%%%%%%%%%%%%%%%%%%% end f3 %%%%%%%%%%%%%%%%%%%%%%%%%%%%%%%%%%
Figure \ref{fig:Phi} shows $\Phi$ calculated for different grain shapes
for various
orientation distributions
[3 values of $\gamma$, and 3 different alignment functions 
$p_{\rm align}(\psi)$], for
suprathermally-rotating grains with
$\bomega$ precessing around $\bB_0$.
We show $\Phi$ for
grain sizes that contribute strongly
to the observed starlight polarization
($\aeff= 0.15\micron$ for $\calP=0$, 
$\aeff= 0.25\micron$ for $\calP=0.5$).

Figure \ref{fig:Phi} shows that $\Phi$ depends
strongly on the grain shape $b/a$, but is
almost independent of both $\gamma$ and the
alignment distribution function $p_{\rm align}$.
This is the case even when 
$p_{\rm align}(\psi)$ is taken to be a delta function 
(as in Figure \ref{fig:Phi})
-- for more realistic (broader)
distributions of alignment angle $\psi$, $\Phi$ would be even less
sensitive to the value of $\gamma$ and the distribution of alignments.

If the MPFA is a good approximation (i.e., if
$\frac{1}{2}\langle C_{{\rm ext},y}-C_{{\rm ext},x}\rangle \approx 
\Cpolext^\MPFA(\lambda)\falign\sin^2\gamma$), 
then
\beq
\Phi(\aeff,b/a,p_{\rm align},\gamma) \approx \Phi^\MPFA(\aeff,b/a)
~,
\eeq
where
\beq \label{eq:PEI}
\Phi^\MPFA \equiv 
\int_{\lambda_1}^{\lambda_2} 
\frac{\Cpolext^\MPFA(b/a,\aeff,\lambda)}{V} d\lambda
\equiv
\int_{\lambda_1}^{\lambda_2} \frac{3}{4}
\frac{\Qpolext^\MPFA(b/a,\aeff,\lambda)}{\aeff} d\lambda
~,
\eeq
where 
$\Cpolext^\MPFA$ is defined in 
Eqs.\ (\ref{eq:CpolMPFA oblate},\ref{eq:CpolMPFA prolate}).

$\Phi^\MPFA$ is plotted vs.\ $b/a$ in Figure \ref{fig:Phi}.
For oblate shapes, $\Phi^\MPFA$ is identical to the
case $\psi=0$ and $\gamma=90^\circ$, but $\Phi^\MPFA$ is in excellent
agreement with $\Phi$ calculated for other cases as well, for both prolate
and oblate shapes.
The upper panels of Fig.\ \ref{fig:Phi} show the
fractional difference between $\Phi$ and $\Phi^\MPFA$ for different
alignment cases.  We see that $\Phi^\MPFA$
approximates the actual $\Phi$ to within a few percent.
We conclude that
the MPFA is an excellent approximation for computing the
polarization efficiency integral for partially-aligned spinning grains.
Thus, for purposes of discussing the polarization efficiency integral
$\Phi$, we do not need to average $C_{{\rm ext},x}$ and 
$C_{{\rm ext},y}$ over
the actual distribution of grain orientations -- 
we can simply take $\Phi \approx \Phi^\MPFA$.

$\Phi^\MPFA(\aeff,b/a)$ is a measure of the polarizing
efficiency for grains of a specified size and shape.
We will show below how 
the observed starlight polarization
can be used to constrain $\Phi^\MPFA$,
thereby constraining the properties of the grains responsible for
polarization of starlight.

% sec 6
\section{\label{sec:polint}
         Starlight Polarization Integral $\Piobs$}

Models to reproduce the extinction and polarization require specifying the
shape of the grains, the grain size distribution $dn_{\rm gr}/d\aeff$, 
and $\falign(\aeff)$,
the fractional alignment 
of grains of size $\aeff$ with the local magnetic field direction.
Suppose the grains to be spheroids with axial ratio $b/a$.
Using the MPFA (Eq.\ \ref{eq:MPFA}),
the polarization is
\beq \label{eq:polmod}
p(\lambda) \approx N_\Ha \int d\aeff 
\left(\frac{1}{\nH} \frac{dn_{\rm gr}}{d\aeff}\right) 
\Cpolext^\MPFA(b/a,\aeff,\lambda)
~\falign(\aeff) \sin^2\gamma
~~~,
\eeq
where $\NH$ is the column density of H nucleons on the sightline.

The strength and wavelength dependence of starlight polarization
have been measured on many sightlines
\citep[e.g.,][]{Serkowski+Mathewson+Ford_1975,
Bagnulo+Cox+Cikota+etal_2017} .
The observed wavelength dependence of the polarization
is quite well described by
the empirical fitting function found by
\citet{Serkowski_1973},  
\beq \label{eq:serkowski}
p(\lambda) \approx p_{\rm max} 
\exp\left[-K\left(\ln(\lambda/\lambdap)\right)^2\right]
~~~.
\eeq
Eq.\ (\ref{eq:serkowski}), referred to as the ``Serkowski law,''
provides a good empirical description of observed starlight polarization from
$2.2\micron$ \citep{Whittet+Martin+Hough+etal_1992}
to wavelengths as short as $0.15\micron$
\citep{Martin+Clayton+Wolff_1999}.

\citet{Serkowski+Mathewson+Ford_1975} suggested
$\lambdap \approx 0.55\micron$ and $K\approx 1.15$ as typical.
Sightline-to-sightline variations in both $\lambdap$ and $K$ are seen,
and are found to be correlated.
\citet{Whittet+Martin+Hough+etal_1992} find
\beq \label{eq:K vs lambdap}
K \approx 0.01 + 1.66 (\lambdap/\micron)
~,
\eeq
although \citet{Bagnulo+Cox+Cikota+etal_2017} report
deviations from this relation.
\citet{Martin+Clayton+Wolff_1999} showed that an improved fit to the
visible-UV polarization was obtained with
\beq
K \approx -0.59 + 2.56 (\lambdap/\micron)
~.
\eeq
As a compromise between the UV and IR, \citet{Whittet_2003}
recommends
\beq
K \approx -0.29 + 2.11 (\lambdap/\micron)
~.
\eeq
At longer wavelengths (3.5$\micron$, 4.8$\micron$) the Serkowski law
(Eq.\ \ref{eq:serkowski}) appears to underestimate the polarization,
and a power-law dependence has been suggested 
\citep{Martin+Whittet_1990}; at even longer wavelengths
there is a prominent polarization feature near $10\micron$ produced
by the Si-O absorption resonance in silicates
\citep{Dyck+Capps+Forrest+Gillett_1973,Smith+Wright+Aitken+etal_2000}.
However, 
the Serkowski law (\ref{eq:serkowski})
provides a generally good fit to observed
polarization from the shortest observed wavelengths ($0.15\micron$)
to $\sim$$2.5\micron$
\citep[see discussion in][]{Hensley+Draine_2021a} 
and we will use it here for that wavelength range.
We will consider $\lambdap=0.55\micron$ and $K=0.87$ as a representative
example (see Table \ref{tab:tab1}).

Observations of starlight polarization 
\citep{Serkowski+Mathewson+Ford_1975,
Bagnulo+Cox+Cikota+etal_2017}
found 
\beq \label{eq:Pmax/E(B-V) opt}
p_{\rm max}\ltsim 0.090 \frac{E(B-V)}{\rm mag}
\eeq
for the sightlines that have been sampled; given that $\ltsim10^2$
sightlines have accurate measurements of $p_{\rm max}/E(B-V)$, 
a small fraction
of sightlines may exceed the limit in Eq.\ (\ref{eq:Pmax/E(B-V) opt}).
From observations of polarized submm emission,
\citet{Planck_2018_XII} 
recommend
\beq \label{eq:Pmax/E(B-V) adopt}
p_{\rm max}\ltsim 0.130 \frac{E(B-V)}{\rm mag}
\eeq
as a more realistic upper limit, and
\citet{Panopoulou+Hensley+Skalidis+etal_2019} found sightlines with
$p_{\rm max}\approx 0.13 E(B-V)/{\rm mag}$.
It is reasonable to suppose that the highest values of
$p_{\rm max}/E(B-V)$ correspond to sightlines where
$\sin^2\!\gamma\approx 1$.  Thus we take
\beq \label{eq:pmax propto E(B-V)}
p_{\rm max}\approx 0.130 \sin^2\!\gamma\, \frac{E(B-V)}{\rm mag} 
~~~.
\eeq

We will see below that the starlight polarization integrated over
wavelength provides a very useful constraint on the population
of aligned grains.
We define the {\it starlight polarization integral} for a sightline
\beq \label{eq:polint}
\Piobs \equiv \int_{\lambda_1}^{\lambda_2} p(\lambda)d\lambda
~~~,
\eeq
with $\lambda_1$ and $\lambda_2$ chosen to
capture the polarization peak:
$\lambda_1\ltsim \lambdap/2$, 
and $\lambda_2\gtsim 2\lambdap$.
If the observed polarization is approximated by 
the Serkowski law (\ref{eq:serkowski}), the polarization integral becomes
\beqa
\Piobs
%\int_{\lambda_1}^{\lambda_2} p(\lambda)d\lambda
&\approx& p_{\rm max} \times \lambdap 
\frac{\sqrt{\pi}\, e^{1/4K}}{2\sqrt{K}}
\left[{\rm erf}(s_1)+{\rm erf}(s_2)\right]
\\
s_1 &\equiv& \sqrt{K}\left[\ln(\lambdap/\lambda_1)+\frac{1}{2K}\right]
\\
s_2 &\equiv& \sqrt{K}\left[\ln(\lambda_2/\lambdap)-\frac{1}{2K}\right]
~~~,
\eeqa
where ${\rm erf}(s)\equiv (2/\sqrt{\pi})\int_0^s e^{-x^2}dx$
is the usual error function.
We set $\lambda_1=0.15\micron$ and $\lambda_2=2.5\micron$, so that
we only use the Serkowski law at wavelengths where it has been
confirmed to be applicable.

%%%%%%%%%%%%%%%%%%%%%%%%%%%%%%% tab 1 %%%%%%%%%%%%%%%%%%%%%%%%%%%%%%%%%%%%%%%%
\begin{table}[ht]
\footnotesize
\begin{center}
\caption{\label{tab:tab1}}
\begin{tabular}{c c c c}
\multicolumn{4}{c}{Observed Starlight Polarization Integral $\Piobs$$^{\,a}$}\\
\hline
$\lambdap$ & $K$ & $\Piobs/p_{\rm max}$ \\
($\micron$)&          & ($\micron$) & reference \\
\hline
0.55 & 1.15 & 1.07 & \citet{Serkowski+Mathewson+Ford_1975}\\
\hline
0.45 & 0.76 & 1.13 & \citet{Whittet+Martin+Hough+etal_1992} \\
0.50 & 0.84 & 1.17 & ''     \\
0.55 & 0.92 & 1.20 & ''     \\
0.60 & 1.01 & 1.23 & ''     \\
0.65 & 1.09 & 1.25 & ''     \\
%0.70 & 1.17 & 1.28 & ''     \\
\hline
0.45 & 0.56 & 1.32     & \citet{Martin+Clayton+Wolff_1999} \\
0.50 & 0.69 & 1.29     & '' \\
0.55 & 0.82 & 1.27     & '' \\
0.60 & 0.95 & 1.26     & '' \\
0.65 & 1.07 & 1.27     & '' \\
\hline
0.45 & 0.66 & 1.22  & \citet{Whittet_2003} \\
0.50 & 0.77 & 1.22  & '' \\
0.55 & 0.87 & 1.23  & '' \\
0.60 & 0.98 & 1.25  & '' \\
0.65 & 1.08 & 1.26  & '' \\
\hline
0.55 & 0.87 & {\bf 1.23} & {\bf representative} \\
\hline
\multicolumn{4}{l}{$^a$ for $\lambda_1=0.15\micron$, $\lambda_2=2.5\micron$}
\end{tabular}
\end{center}
\end{table}
%%%%%%%%%%%%%%%%%%%%%%%%%%%%%% end tab 1 %%%%%%%%%%%%%%%%%%%%%%%%%%%%%%%%%%%%%%%

Values of $\Piobs/p_{\rm max}$ are given in Table \ref{tab:tab1} for
various $\lambdap$ and $K$.
It is noteworthy that $\Piobs/p_{\rm max}$ is not very sensitive
to the precise values of $\lambdap$ and $K$.
We take $\Piobs/p_{\rm max}\approx 1.23\micron$ as a
representative value for the diffuse ISM.

% sec 6
%\section{\label{sec:Phi}
%         Starlight Polarization Efficiency Integral $\Phi$}

\section{\label{sec:lambda_peff}
         Characteristic Size of the Aligned Grains}

The dust
in the ISM must have
shapes and sizes such that the population of aligned grains reproduces
the 
the Serkowski law (\ref{eq:serkowski}); this requires a distribution of
grain sizes, but the most important grains will be the ones contributing
to the polarization at wavelengths 
near the peak at $\lambdap\approx0.55\micron$.

We define
an effective wavelength $\lambda_{\rm p,eff}(\aeff,b/a)$ for the polarization
contribution by grains of a given size $\aeff$ and shape $b/a$:
\beq
\lambda_{\rm p,eff}
\equiv
\exp\left[
\frac
{\int_{\lambda_1}^{\lambda_2} (\ln\lambda)\, 
\Qpolext^\MPFA (b/a,\aeff,\lambda)\, d\ln\lambda}
{\int_{\lambda_1}^{\lambda_2} 
\Qpolext^\MPFA (b/a,\aeff,\lambda)\, d\ln\lambda}
\right]
~~~.~~
\eeq
%%%%%%%%%%%%%%%%%%%%%%%%%%%%%%%%%%%% f4 %%%%%%%%%%%%%%%%%%%%%%%%%%%%%%%%%%%%%%
\begin{figure}[ht]
\begin{center}
\includegraphics[angle=0,width=8.0cm,
                 clip=true,trim=0.5cm 5.0cm 0.5cm 2.5cm]%l b r t
{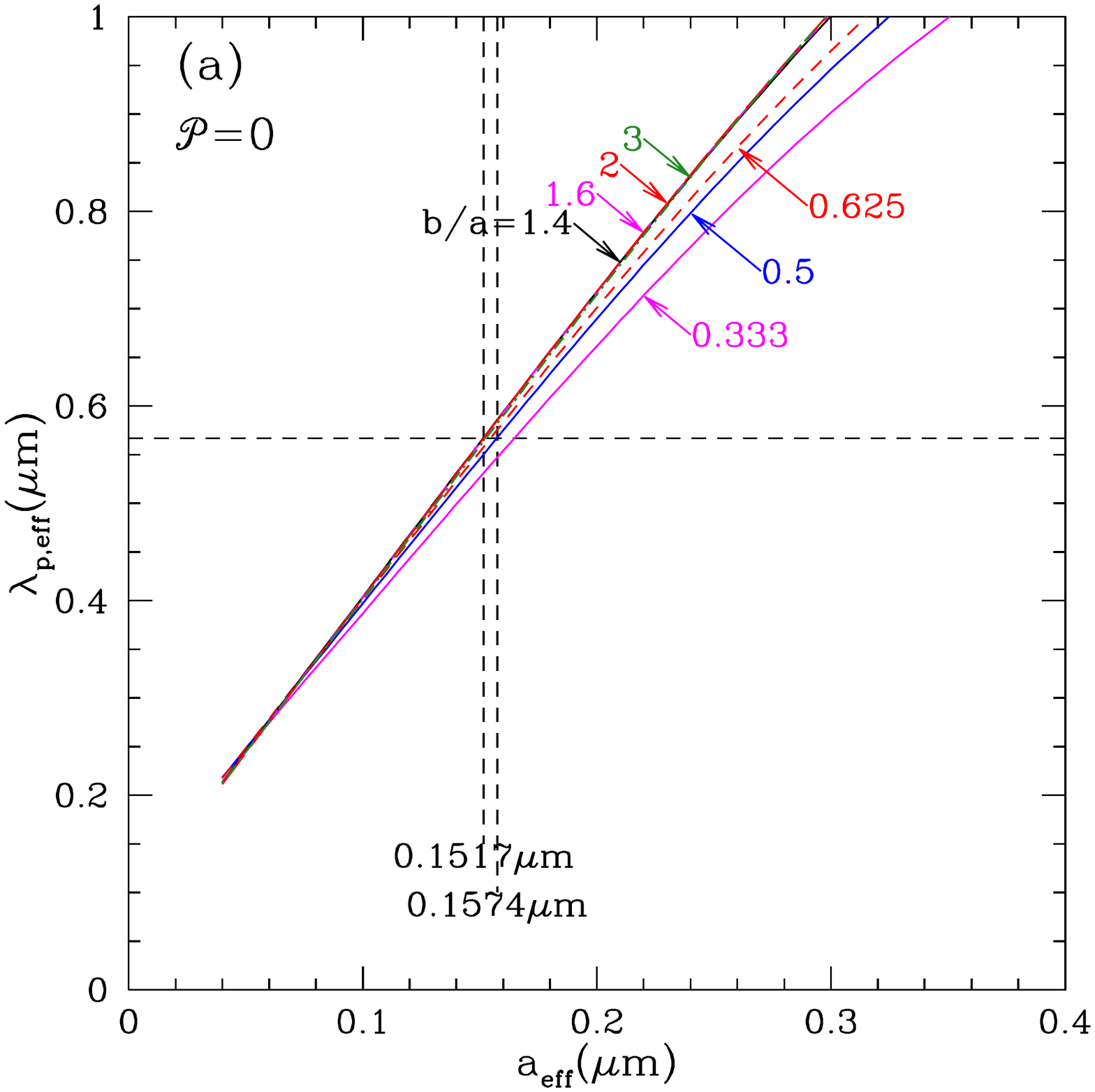}
\includegraphics[angle=0,width=8.0cm,
                 clip=true,trim=0.5cm 5.0cm 0.5cm 2.5cm]%l b r t
{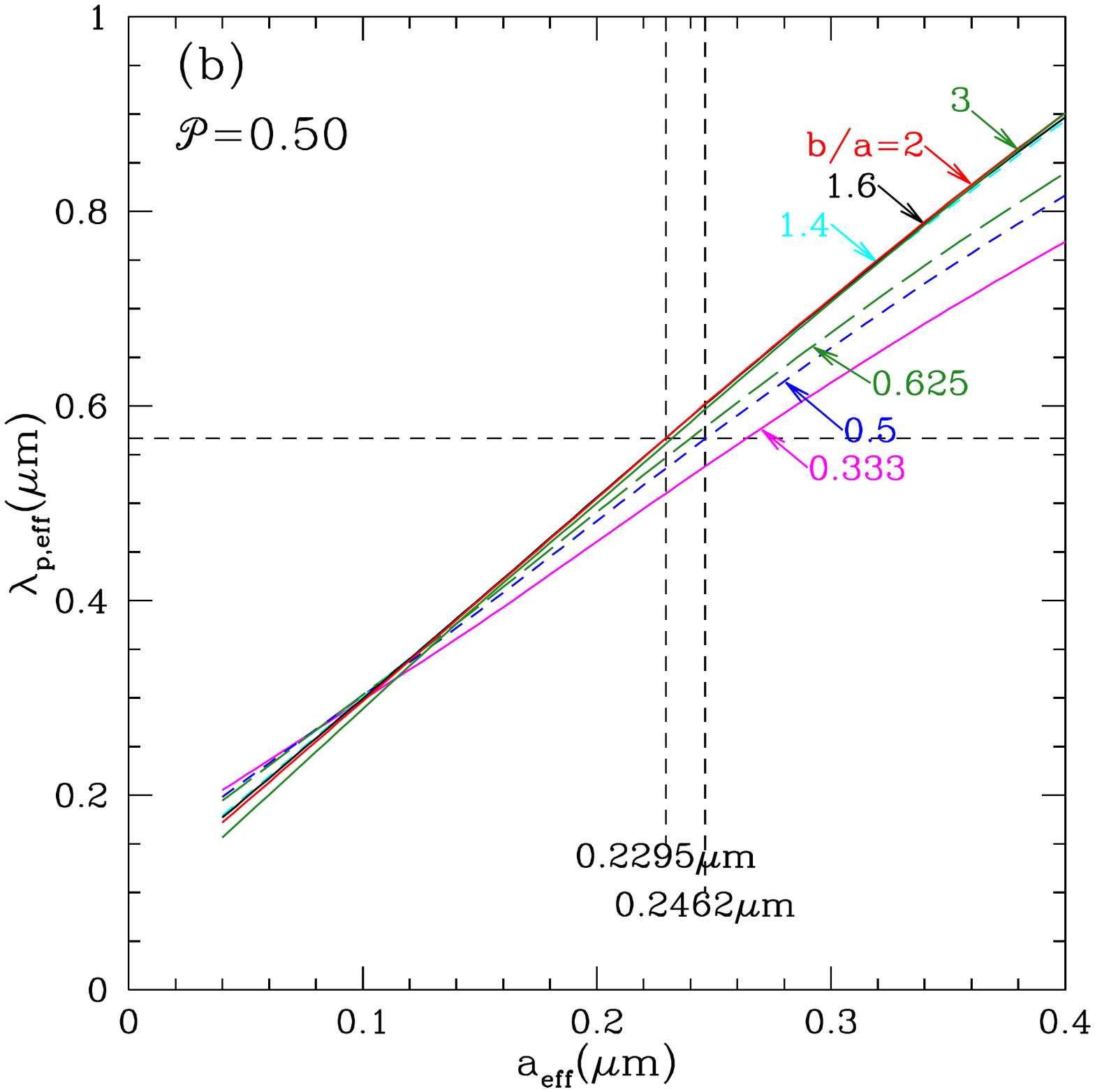}
\caption{\footnotesize \label{fig:lambda_peff}
Effective wavelength $\lambda_{\rm p,eff}$
for polarization as a function of $\aeff$, for different axial ratios.
(a) Astrodust grains with no porosity.
(b) Astrodust grains with porosity $\calP=0.5$
The horizontal dashed line shows $\lambda_{\rm p,eff}=0.567\micron$ 
for the Serkowski law
with $\lambdap=0.55\micron$ and $K=0.87$.
\btdnote{uses results from eval\_Phi....}
\btdnote{f4a.pdf,f4b.pdf}
}
\vspace*{-0.3cm}
\end{center}
\end{figure}
%%%%%%%%%%%%%%%%%%%%%%%%%%%% end f4 %%%%%%%%%%%%%%%%%%%%%%%%%%%%%%%%%%%%%%%
We can also calculate the effective wavelength
$\lambda_{\rm p,eff}^\obs$
for the Serkowski law (\ref{eq:serkowski}):
\beqa
\ln(\lambda_{\rm p,eff}^\obs)
&\equiv&
\frac
{\int_{\lambda_1}^{\lambda_2} (\ln\lambda)\, p (\lambda)\, d\ln\lambda}
{\int_{\lambda_1}^{\lambda_2} p (\lambda)\, d\ln\lambda}\\
&=& \ln(\lambdap) + \frac{1}{\sqrt{\pi K}}
\frac{\exp[-K(\ln(\lambdap/\lambda_1))^2]-\exp[-K(\ln(\lambda_2/\lambdap))^2]}
{{\rm erf}[\sqrt{K}\ln(\lambdap/\lambda_1)]+{\rm erf}[\sqrt{K}\ln(\lambda_2/\lambdap)]}
%\\
%&=& \ln(\lambdap) + 0.0305
\\
\lambda_{\rm p,eff}^\obs &=& 0.567\micron
~~~{\rm for~}\lambda_p=0.55\micron,~ K=0.87
~~~.
\eeqa
%for the representative case $\lambdap=0.55\micron$, $K=0.87$
%(for $\lambda_1=0.15\micron$, 
%$\lambda_2=2.5\micron$).
Thus, the grains that dominate the starlight polarization should have sizes
$\aeff$ such that $\lambda_{\rm p,eff}\approx 0.57\micron$.

%%%%%%%%%%%%%%%%%%%%%%%%%%%  f5 %%%%%%%%%%%%%%%%%%%%%%%%%%%%%%%%%%%%%%%%%%%
\begin{figure}[ht]
\begin{center}
\includegraphics[angle=0,width=8.0cm,
                 clip=true,trim=0.5cm 5.0cm 0.5cm 2.5cm]%l b r t
{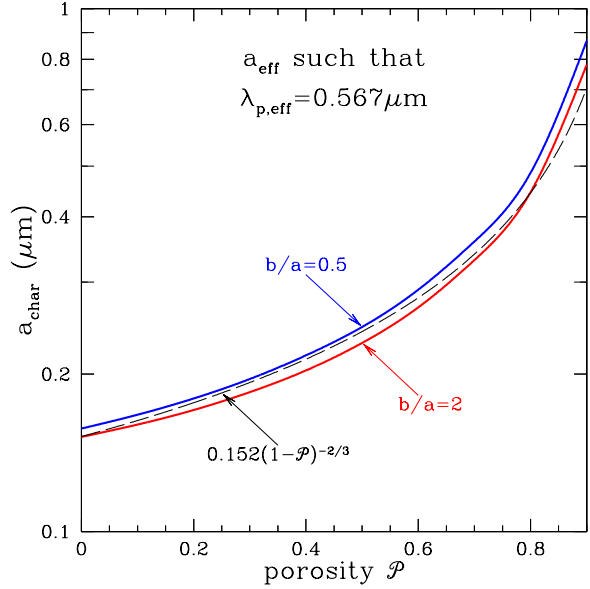}
\caption{\footnotesize \label{fig:aeff vs P}
Characteristic grain size $a_{\rm char}$
such that $\lambda_{p,{\rm eff}}=0.567\micron$
as a function of porosity $\poro$, for oblate ($b/a=2$) and prolate
($b/a=0.5$) spheroids.
Also shown (dashed curve)
is fitting function $a_{\rm char}=0.152\micron(1-\poro)^{-2/3}$.
\btdnote{f5.pdf}
}
\vspace*{-0.3cm}
\end{center}
\end{figure}
%%%%%%%%%%%%%%%%%%%%%%%%%%% end f5 %%%%%%%%%%%%%%%%%%%%%%%%%%%%%%%%%%%%%%%%
$\Qpolext^\MPFA(\lambda)$ and $\lambda_{\rm p,eff}$ depend on the
adopted dielectric function and on grain size and shape.
Figure \ref{fig:lambda_peff}
shows $\lambda_{\rm p,eff}$ as a function of $\aeff$ for astrodust grains with
low and high porosity.
The observed polarization, peaking near $\lambdap\approx0.55\micron$, is
evidently dominated by grains with sizes $\aeff$ close to 
a characteristic grain size 
$a_{\rm char}$, defined to be the size for which $\lambda_{\rm p,eff}(a)
\approx \lambda_{\rm p,eff}^\obs=0.567\micron$.
We see that $a_{\rm char}\approx 0.15\micron$ for $\poro=0$, and
$a_{\rm char}\approx 0.24\micron$ if the porosity $\calP\approx0.5$.
Figure \ref{fig:aeff vs P} shows how $a_{\rm char}$ depends on
$\poro$ -- the numerical results are approximated by the
simple fitting function
\beq \label{eq:achar}
a_{\rm char} \approx \frac{0.152\micron}{(1-\poro)^{2/3}}
~.
\eeq 

% sec 8
\section{\label{sec:falign}
         $\Phi^\MPFA$, Aligned Mass Fraction $\langle \falign\rangle$,
         and Limits on Axial Ratios}

Models to reproduce the observed wavelength-dependent extinction
typically have grain mass distributions peaking around $\aeff\approx 0.25\micron$,
but with appreciable mass in the $a\ltsim0.1\micron$ grains that are required
to reproduce the observed rapid rise in extinction into the UV.
The classic MRN model \citep{Mathis+Rumpl+Nordsieck_1977}
reproduces the extinction using spherical grains 
with $dn/da \propto a^{-3.5}$ for $0.005\micron\leq a\leq 0.25\micron$;
this distribution has $45\%$ of the grain mass
in grains with $a<0.07\micron$.
More recent grain models with more complicated size distributions
\citep[e.g.,][]{Weingartner+Draine_2001a,Zubko+Dwek+Arendt_2004}
have similar fractions of the silicate mass in grain sizes $\aeff<0.07\micron$.
However, models to reproduce the polarization of starlight 
\citep[e.g.,][Hensley \& Draine 2021, in prep.]{Kim+Martin_1995b,
                Draine+Fraisse_2009,
                Siebenmorgen+Voshchinnikov+Bagnulo_2014,
                Fanciullo+Guillet+Boulanger+Jones_2017,
                Guillet+Fanciullo+Verstraete+etal_2018}
require that grains smaller than $\ltsim 0.07\micron$ 
have minimal alignment --
if these grains were aligned the polarization would exceed
the observed low polarization in the UV
\citep{Martin+Clayton+Wolff_1999}.
If we estimate $>30$\% of the dust mass to be in grains
that are {\it not} aligned, then the mass-averaged alignment efficiency
\beq \label{eq:Phisbgmin}
\langle \falign\rangle 
< 0.70
~~~.
\eeq

\begin{figure}[h]
\begin{center}
\includegraphics[angle=0,width=5.8cm,
                 clip=true,trim=0.5cm 0.5cm 0.5cm 0.5cm]%l b r t
{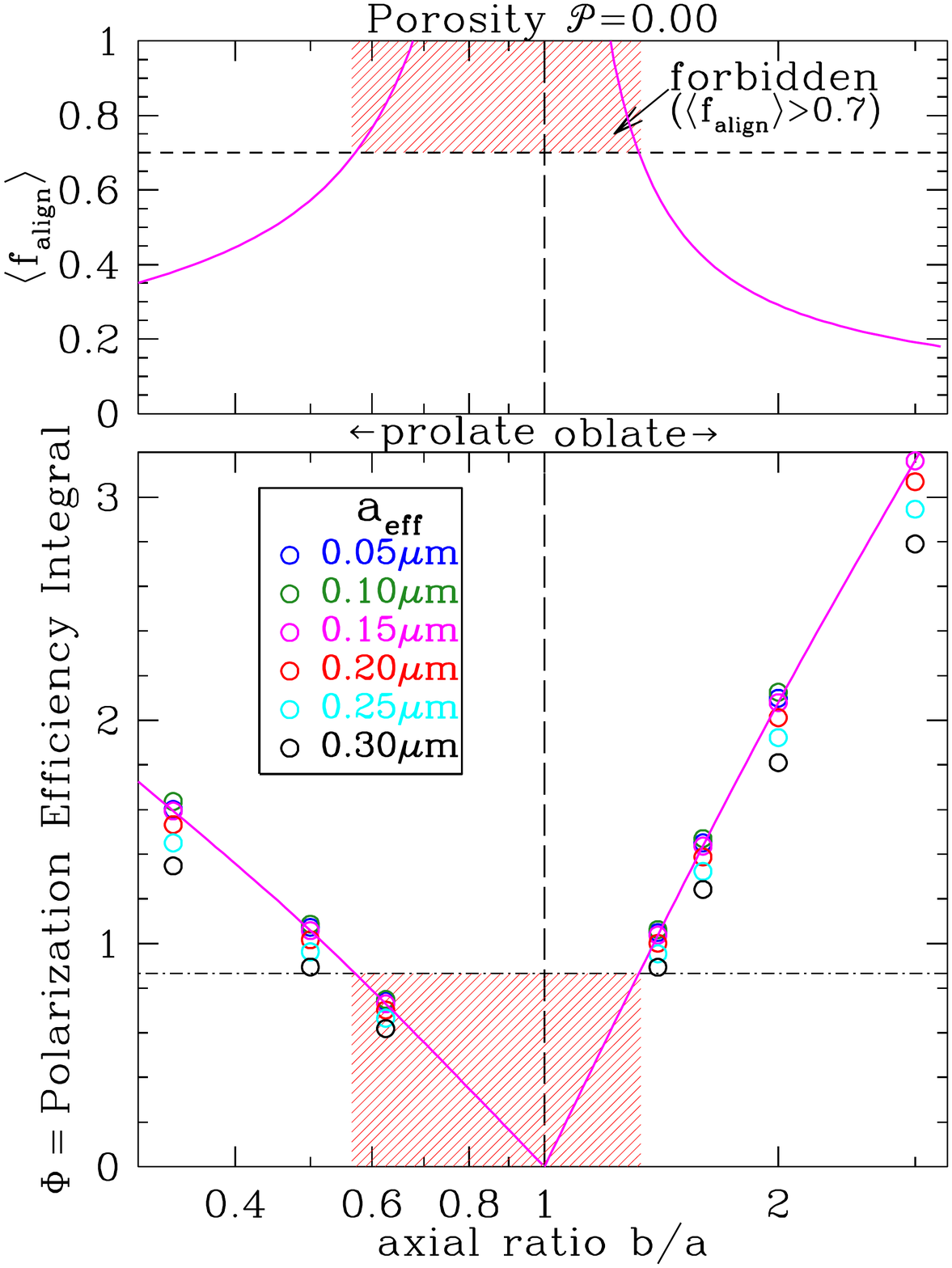}
\includegraphics[angle=0,width=5.8cm,
                 clip=true,trim=0.5cm 0.5cm 0.5cm 0.5cm]%l b r t
{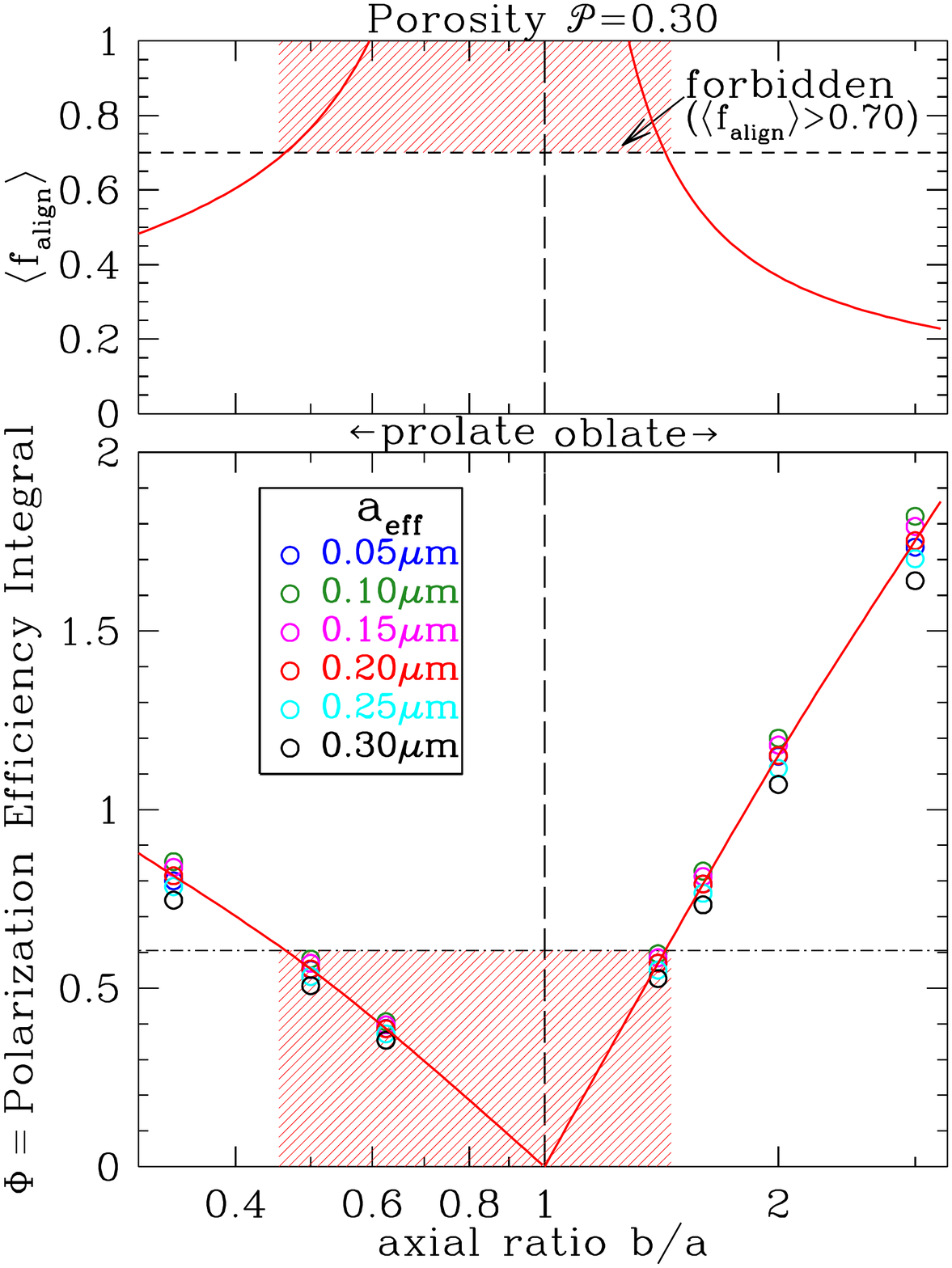}
\includegraphics[angle=0,width=5.8cm,
                 clip=true,trim=0.5cm 0.5cm 0.5cm 0.5cm]%l b r t
{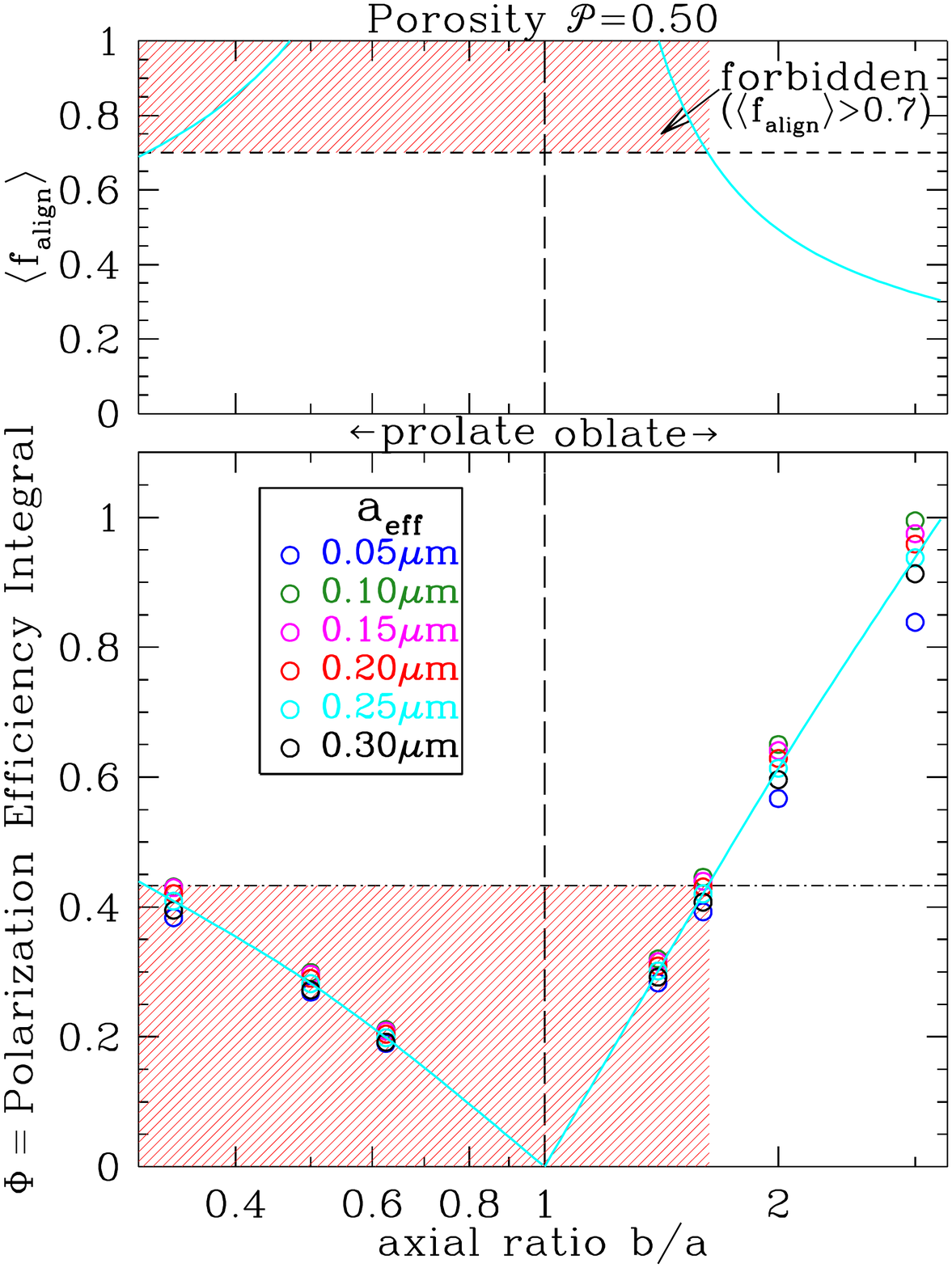}
\caption{\label{fig:Phisbg vs ba} \footnotesize
         Lower panels: Starlight polarization efficiency integral $\Phisbg$ for 
         ``astrodust'' spheroids
         as a function of axial ratio $b/a$, 
         for 6 grain sizes
         and 3 different
         porosities $\calP$.
         Upper panels: $\langle\falign\rangle$, 
         the aligned mass fraction
         of the astrodust grains.  
         Values of $\langle\falign\rangle >0.7$ are forbidden
         (shaded regions).
         \btdnote{f6a.pdf, f6b.pdf, f6c.pdf}
         }
\vspace*{-0.3cm}
\end{center}
\end{figure}

Figure \ref{fig:Phisbg vs ba} (lower panels) shows 
the starlight polarization efficiency integral $\Phisbg$ calculated for
astrodust grains, as a function of
axial ratio $b/a$, for selected sizes $\aeff$, 
for three cases:
zero
porosity ($\calP=0$),
moderate porosity ($\calP=0.30$),
and high porosity
($\calP=0.50$).
%For each $\calP$, note the insensitivity of $\Phisbg$ to $\aeff$.
%thus justifying the use of $\Phi(b/a,\achar)$ in Eq.\ (\ref{eq:Phi=const}).
Increasing porosity results in a significant decrease
in $\Phisbg$, because, per unit grain mass, the more porous grains are less
effective polarizers.

%%%%%%%%%%%%%%%%%%%%%%%%%%%%%%%%% f6 %%%%%%%%%%%%%%%%%%%%%%%%%%%%%%%
%%%%%%%%%%%%%%%%%%%%%%%%%%%%%%%%%%%%%% end f6 %%%%%%%%%%%%%%%%%%%%%%%%%%%%
For the grain sizes 
required to
reproduce the observed polarization of starlight, the starlight
polarization efficiency factor $\Phisbg(b/a,\aeff)$ is seen to be
almost independent of $\aeff$.
The curves in Figure \ref{fig:Phisbg vs ba} are for
$a_{\rm eff}=a_{\rm char}$, with $a_{\rm char}$ from Eq.\ (\ref{eq:achar}).

For an assumed grain shape,
the total volume of the
aligned grains can be estimated from
the (observed) starlight polarization integral $\Piobs$
and the (theoretical) polarization efficiency integral $\Phi(b/a)$
without need to
solve for the actual size distribution of aligned grains.
Integrating
Eq.\ (\ref{eq:polmod}) over wavelength, we obtain
\beqa \label{eq:relate Phi to polint}
\Piobs &\,=\,& 
\NH \int d\aeff 
\left(\frac{1}{\nH}\frac{dn_{\rm gr}}{d\aeff}\right)
\frac{4}{3}\pi\aeff^3 
~\Phi(b/a,\aeff) ~\falign(\aeff) \sin^2\!\gamma
\\ \label{eq:Phi=const}
&\approx& \NH ~\Phi(b/a,a_{\rm char}) ~\sin^2\!\gamma
\int d\aeff \left(\frac{1}{\nH}\frac{dn_{\rm gr}}{d\aeff}\right)
\frac{4}{3}\pi\aeff^3
~\falign(\aeff)
~,~~ 
\eeqa
where in Eq.\ (\ref{eq:Phi=const}) 
we take $\Phi(\aeff)$ to be approximately constant for
grain sizes $\aeff$ near the characteristic grain size $a_{\rm char}$.
Let $V_{\rm align}$ 
be the volume per H of aligned grains:
\beq \label{eq:integral=integral}
V_{\rm align}
\equiv
\int d\aeff 
\left(\frac{1}{\nH} \frac{dn_{\rm gr}}{d\aeff}\right)
\frac{4}{3}\pi\aeff^3 ~ \falign(\aeff)
~~~.
\eeq
Eq.\ (\ref{eq:Phi=const}) and (\ref{eq:integral=integral}) become
\beq
\Piobs
\approx
\NH \, \Phisbg(b/a,\achar) \,
\sin^2\!\gamma \,
V_{\rm align} 
% p_{\rm max}\lambdap 
% \times
% \frac{\sqrt{\pi} e^{1/4K}}{2\sqrt{K}}
% \left[{\rm erf}(s_1)+{\rm erf}(s_2)\right]
~~~.
\eeq
Thus we can estimate $V_{\rm align}$,
assuming $\Piobs/p_{\rm max}=1.23\micron$ (see Table \ref{tab:tab1}),
Eq.\ (\ref{eq:pmax propto E(B-V)}), and
\beq \label{eq:NH/E(B-V)}
\NH/E(B-V)\approx 8.8\xtimes10^{21}\cm^{-2}{\rm mag}^{-1}
\eeq
\citep{Lenz+Hensley+Dore_2017}:
\beqa
V_{\rm align}
&\,\approx\,&
\frac{p_{\rm max}/E(B-V)}{\NH/E(B-V)}
\times
\frac{\Piobs/p_{\rm max}}{\Phisbg(b/a,\achar)}
\\ \label{eq:Valign}
&\approx& \frac{0.130}{8.8\xtimes10^{21}\cm^{-2}}
 \times \frac{1.23\micron}{\Phisbg(b/a,\achar)}
=
 \frac{1.82\xtimes10^{-27}\cm^3\Ha^{-1}}{\Phisbg(b/a,\achar)}
~,
\eeqa
for the representative case $\lambdap=0.55\micron$ and $K=0.87$ 
(see Table \ref{tab:tab1}).
For comparison, the total volume of astrodust grains per H nucleon
in the diffuse high-latitude ISM is estimated to be
\beq \label{eq:VAd}
V_{\rm Ad}=3.0\xtimes10^{-27}(1-\calP)^{-1}\cm^3\Ha^{-1}
\eeq
\citep{Draine+Hensley_2021a}.
The mass-weighted alignment efficiency is then
\beq \label{eq:falignsbg}
\langle \falign\rangle \equiv 
\frac{V_{\rm align}}{V_{\rm Ad}} \approx 
\frac{0.61(1-\calP)}{\Phisbg(b/a,\achar)}
~~~.
\eeq

%%%%%%%%%%%%%%%%%%%%%%%%%%%%%%%%%% f7 %%%%%%%%%%%%%%%%%%%%%%%%%%%%%%%%
\begin{figure}[t]
\begin{center}
\includegraphics[angle=270,width=10.0cm,
                 clip=true,trim=0.5cm 0.5cm 0.5cm 0.5cm]%l b r t
{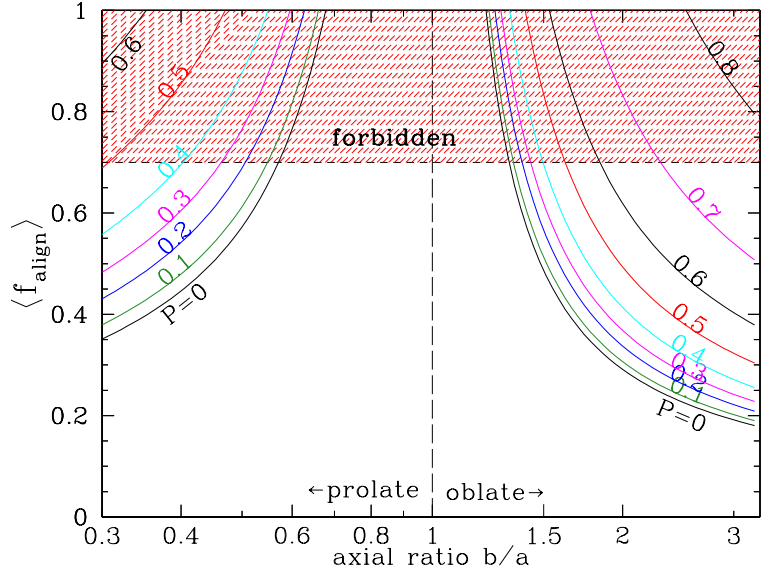}
\caption{\label{fig:falign} \footnotesize
         Mass-weighted aligned fraction of astrodust grains, 
         $\langle \falign\rangle$
         (see Eq.\ \ref{eq:falignsbg}), as a function of axial ratio $b/a$.
         The region $\langle \falign\rangle \gtsim 0.7$ 
         (shaded in red) is
         forbidden by
         low levels of polarization in the ultraviolet (see text).
         \btdnote{f7.pdf}
         }
\vspace*{-0.3cm}
\end{center}
\end{figure}
%%%%%%%%%%%%%%%%%%%%%%%%%%%%%%%%% end f7 %%%%%%%%%%%%%%%%%%%%%%%%%%%%%%%%%%

%This is the shaded region in Figure \ref{fig:falign}.

The upper panels in Figure \ref{fig:Phisbg vs ba} show
$\langle\falign\rangle$ required to account for the
observed polarization of starlight on sightlines with the largest
$p_{\rm max}/E(B-V)$ [see Eq.\ (\ref{eq:falignsbg})].
The shaded regions in Figure \ref{fig:Phisbg vs ba} correspond to 
$\langle \falign \rangle>0.7$,
which is forbidden (see Eq.\ \ref{eq:Phisbgmin}).
For $\calP=0$, Figure \ref{fig:Phisbg vs ba} shows that
(\ref{eq:Phisbgmin}) requires
prolate grains to have axial ratios $a/b \gtsim 1.8$, and
oblate grains to have axial ratios $b/a \gtsim 1.35$ --
less extreme axial ratios would not produce sufficient starlight
polarization to reproduce observations, even with 70\% of the dust
mass perfectly aligned.
For larger porosities ($\poro=0.3$ and $0.5$),
Figure \ref{fig:Phisbg vs ba} shows that more extreme axial ratios
are required, because porous grains are less efficient polarizers.

Figure \ref{fig:falign} shows $\langle \falign\rangle$ 
as a function
of axial ratio $b/a$, for selected porosities $\poro$ from 0 to 0.8.
As expected, $\langle \falign\rangle$ 
is a decreasing function of
aspect ratio, because more extreme shapes are better polarizers.

If astrodust grains are porous,
the total grain volume $V_{\rm Ad}$ is increased, and smaller
values of $\Phisbg$ are allowed.
However, porous grains are less effective polarizers per unit mass,
and the net effect is to require the astrodust grains to have
more extreme
axial ratios as $\calP$ is increased -- see
Fig.\ \ref{fig:falign}. 

%%%%%%%%%%%%%%%%%%%%%%%%%%%%%%%%%%%%%% f8 %%%%%%%%%%%%%%%%%%%%%%%%%%%%%
\begin{figure}[ht]
\begin{center}
\includegraphics[angle=0,width=10.0cm,
                 clip=true,trim=0.5cm 5.0cm 0.5cm 2.5cm]%l b  t
{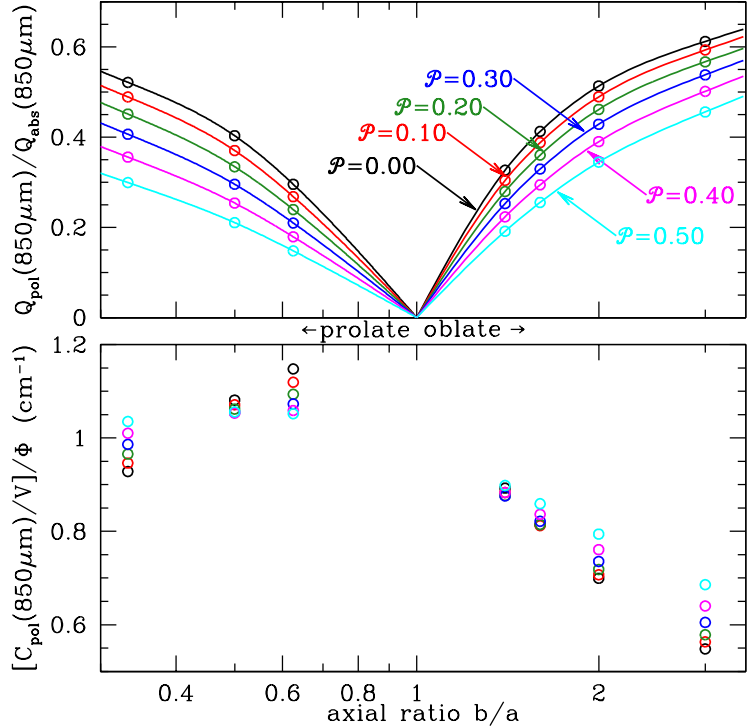}
\caption{\label{fig:Qpol850/Qabs850} \footnotesize
         Upper panel: 
         Ratio of submm ($\lambda=850\micron$) polarization cross section to
         absorption cross section for astrodust spheroids, as a function
         of axial ratio $b/a$.
         Lower panel:
         Ratio of submm polarization cross section per volume
         $\Cpolext/V$ divided by starlight polarization efficiency factor
         $\Phi$ for astrodust grains, as a function of axial ratio $b/a$.
         This measures the ability to polarize in the submm relative
         to the optical.
         %Oblate shapes have lower values of $(\Cpolext/V)/\Phi$ because they
         %have higher values of $\Phi$ (see Fig.\ \ref{fig:Phisbg vs ba}).
         \btdnote{f8.pdf}
         }
\vspace*{-0.3cm}
\end{center}
\end{figure}
%%%%%%%%%%%%%%%%%%%%%%%%%%%%%%%%%%%% end f8 %%%%%%%%%%%%%%%%%%%%%%%%%%%%%%

%%%%%%%%%%%%%%%%%%%%%%%%%%%%%%%%% f9 %%%%%%%%%%%%%%%%%%%%%%%%%%%%%%%
\begin{figure}[ht]
\begin{center}
\includegraphics[angle=0,width=5.8cm,
                 clip=true,trim=0.5cm 5.0cm 0.5cm 2.5cm]%l b r t
{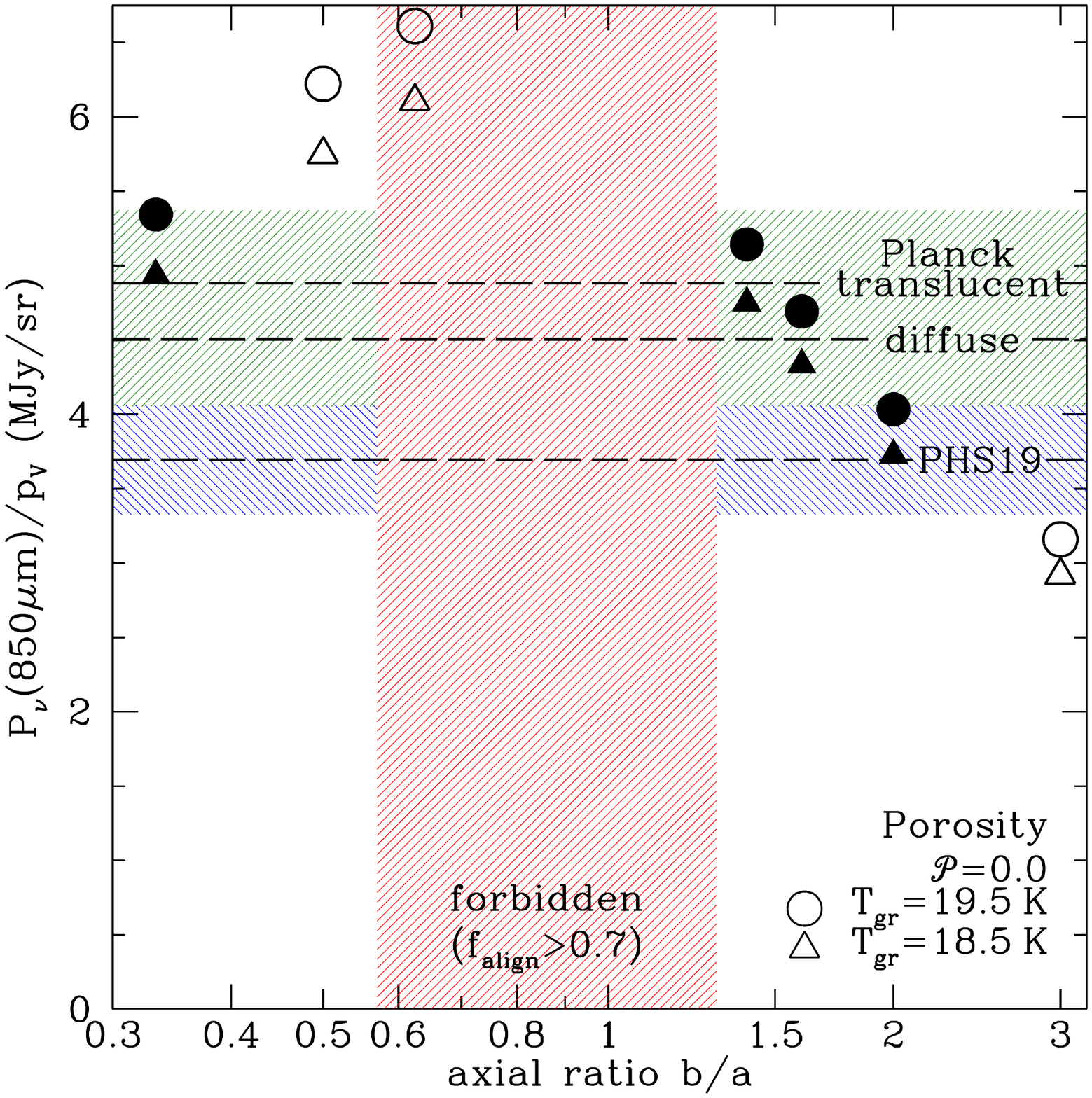}
\includegraphics[angle=0,width=5.8cm,
                 clip=true,trim=0.5cm 5.0cm 0.5cm 2.5cm]%l b r t
{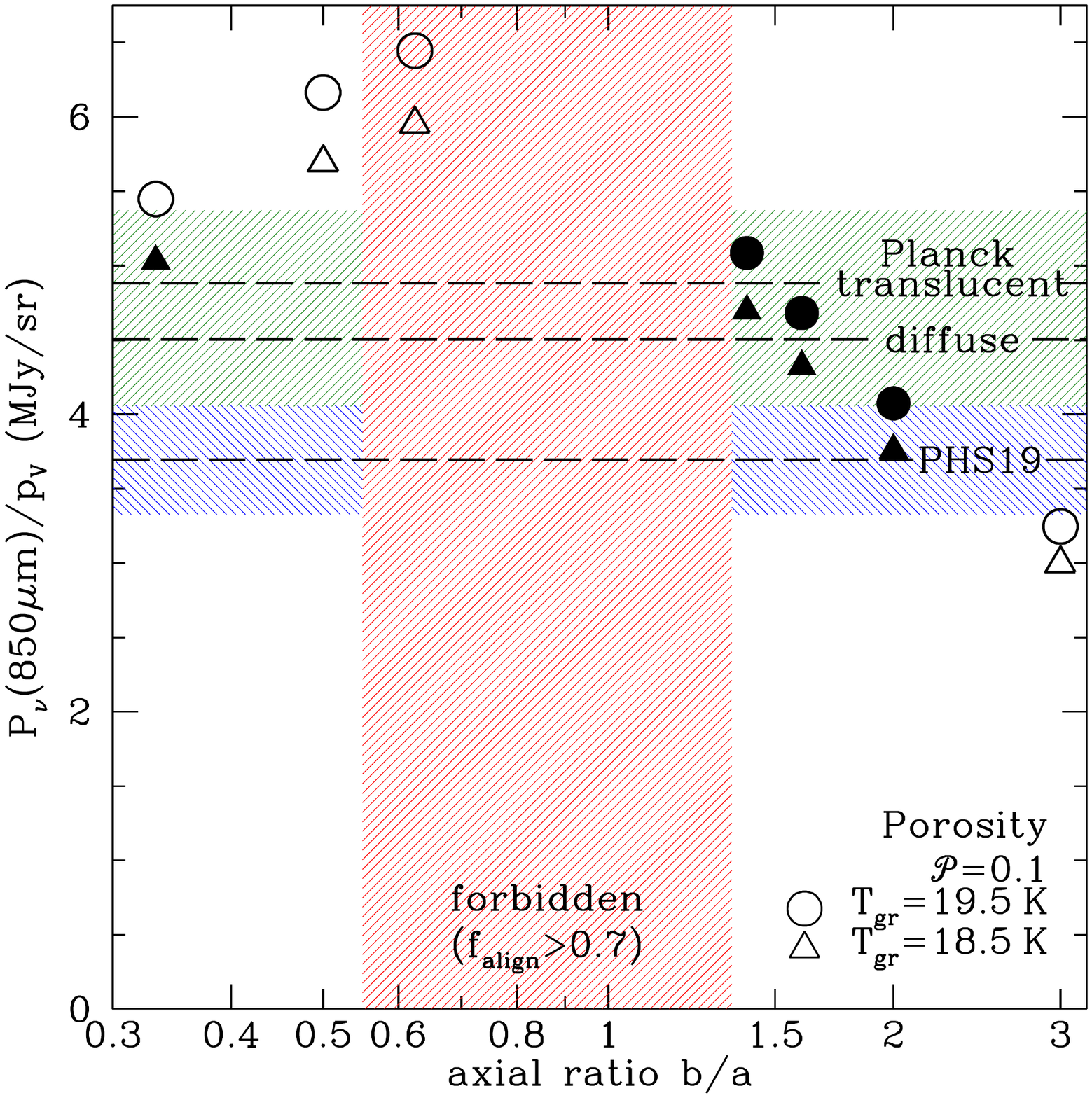}
\includegraphics[angle=0,width=5.8cm,
                 clip=true,trim=0.5cm 5.0cm 0.5cm 2.5cm]%l b r t
{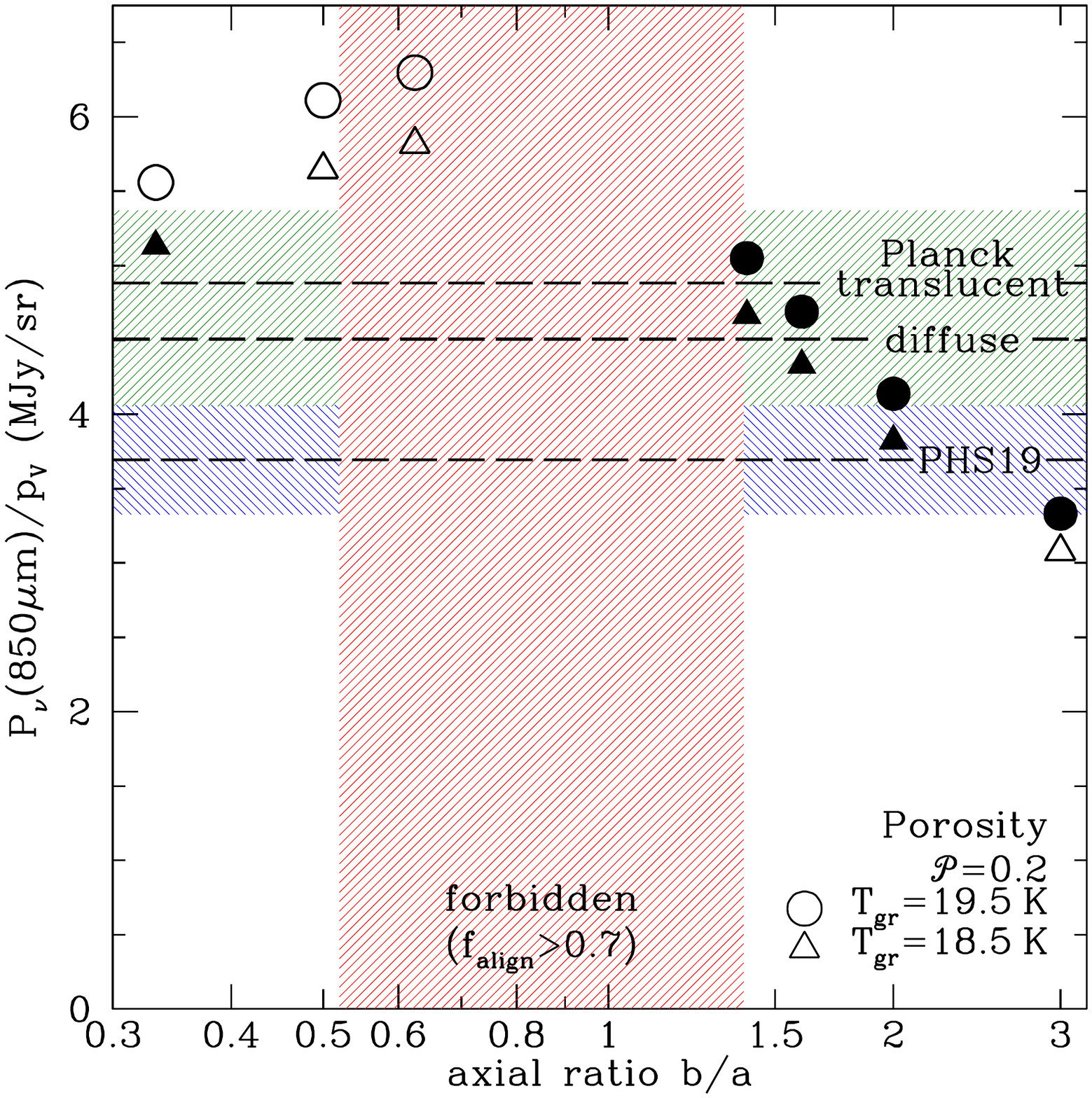}
\includegraphics[angle=0,width=5.8cm,
                 clip=true,trim=0.5cm 5.0cm 0.5cm 2.5cm]%l b r t
{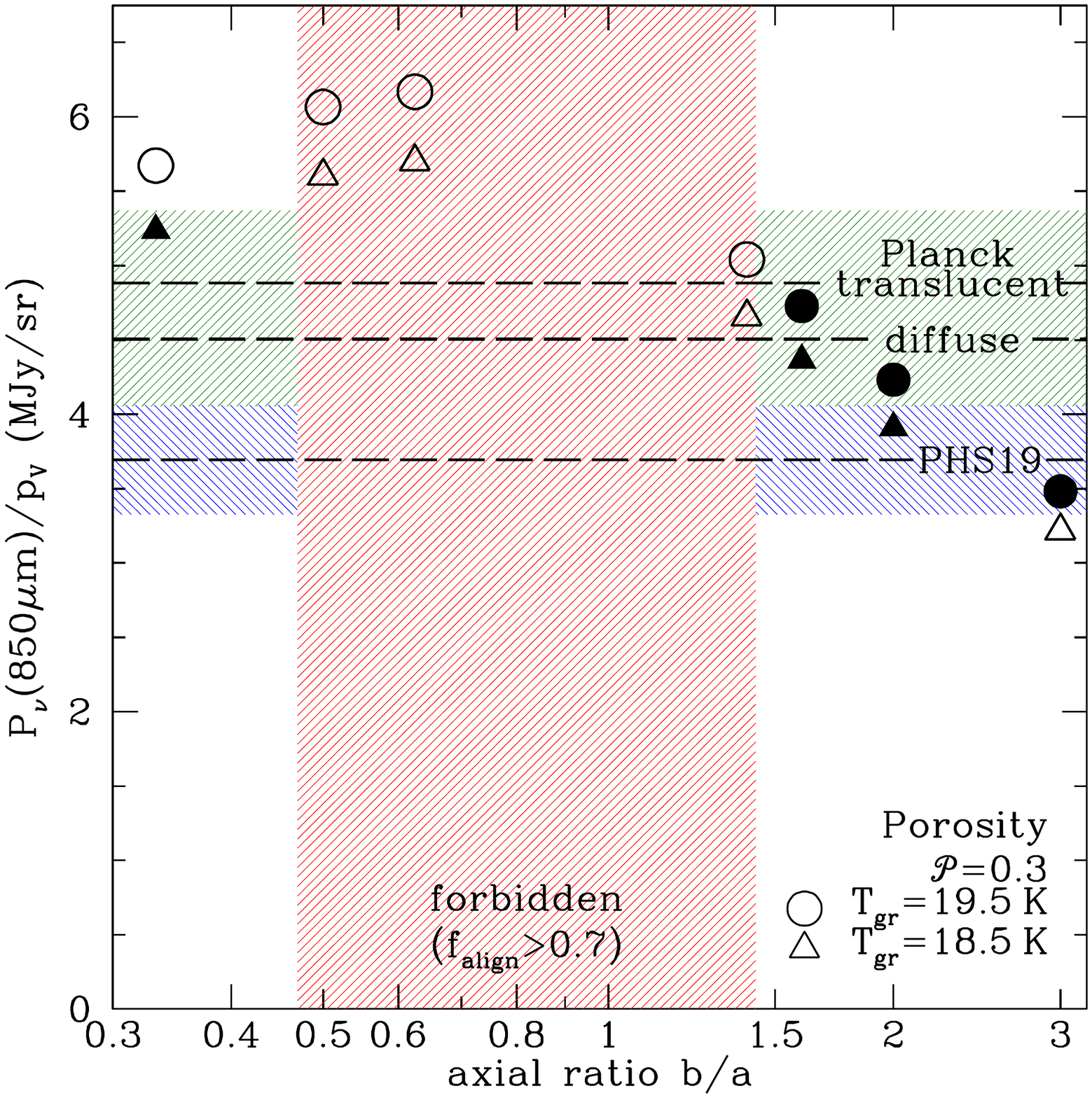}
\includegraphics[angle=0,width=5.8cm,
                 clip=true,trim=0.5cm 5.0cm 0.5cm 2.5cm]%l b r t
{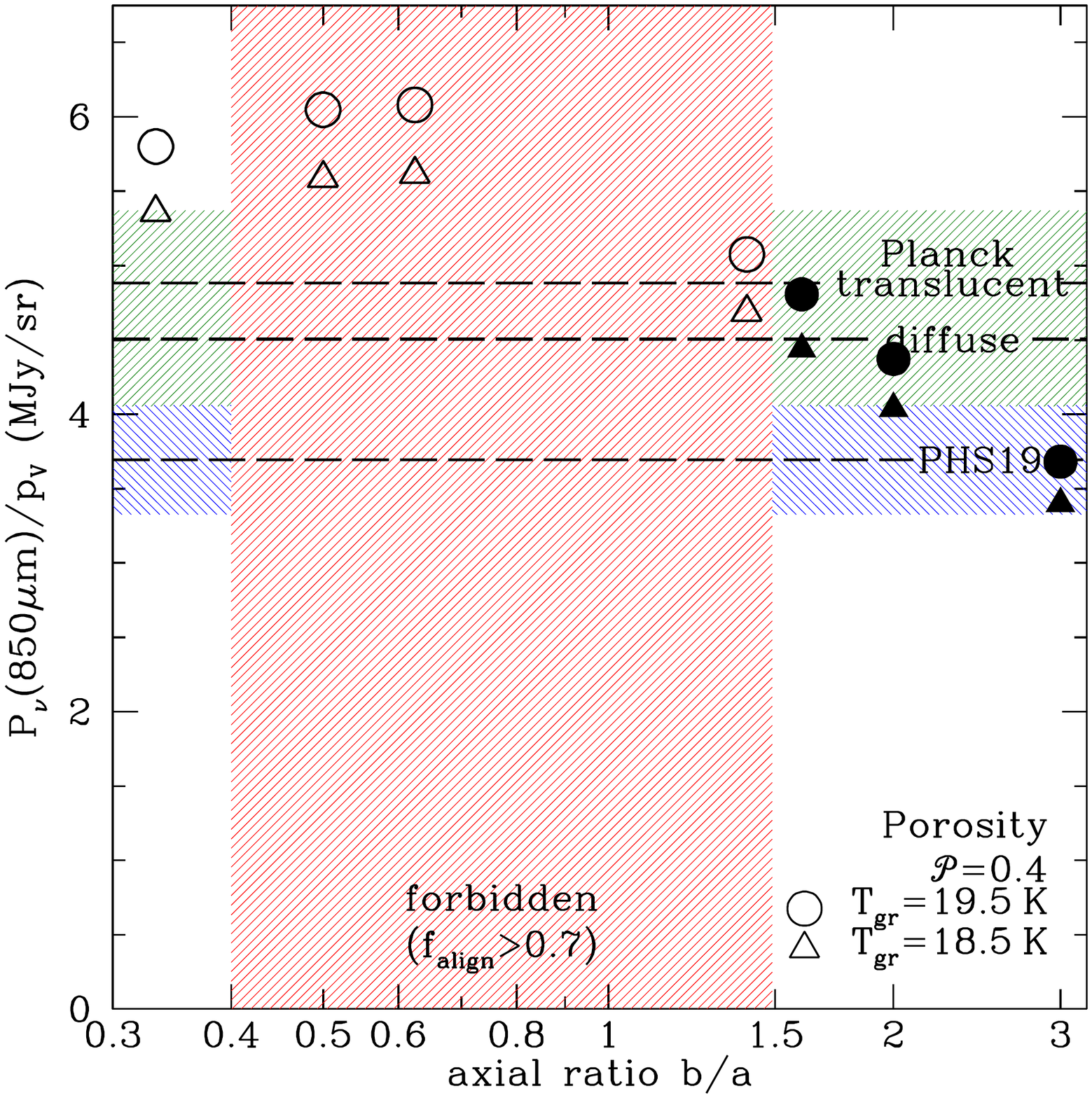}
\includegraphics[angle=0,width=5.8cm,
                 clip=true,trim=0.5cm 5.0cm 0.5cm 2.5cm]%l b r t
{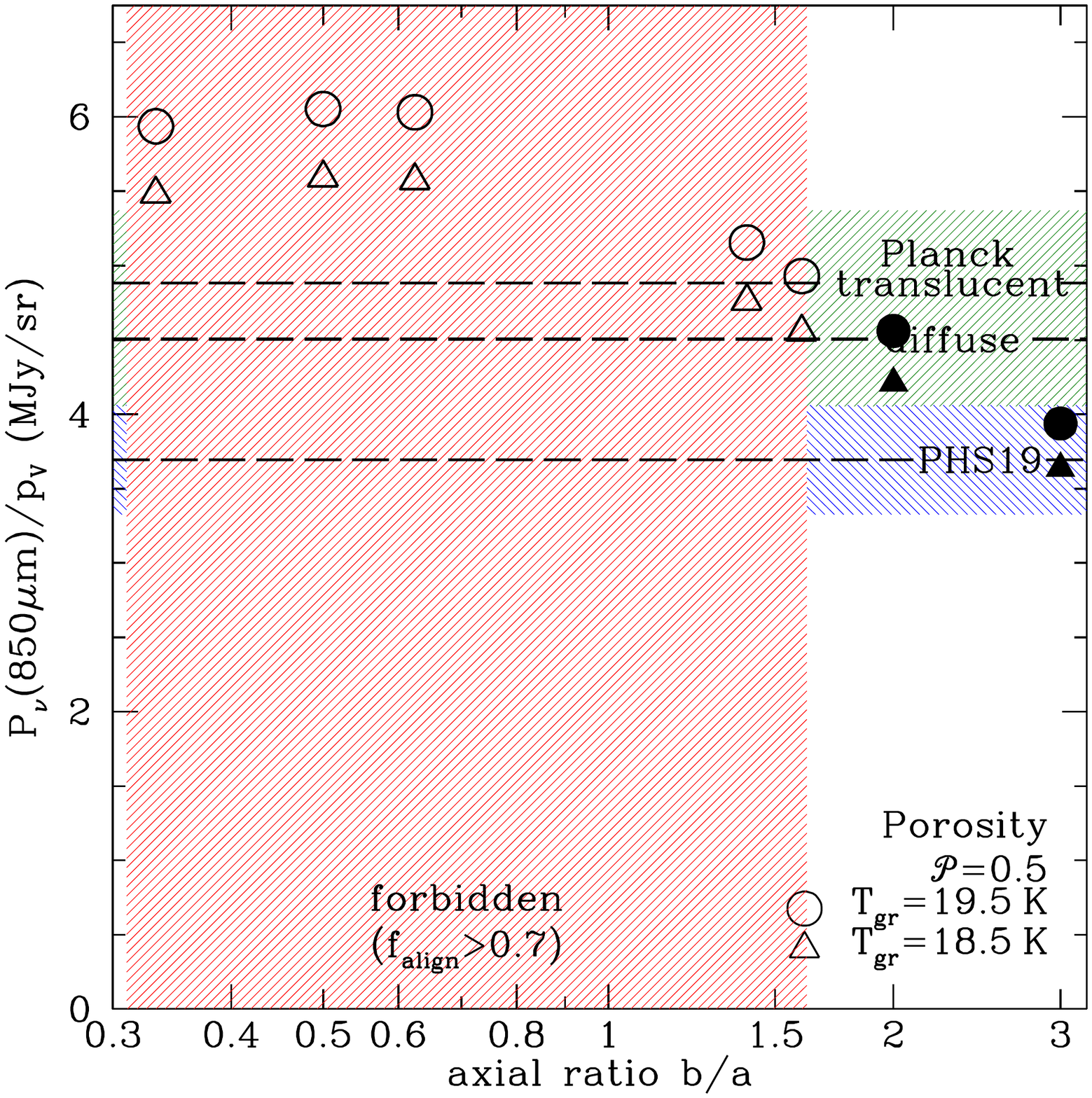}
\caption{\label{fig:P_nu/p_V vs ba} \footnotesize
         $P_\nu(850\micron)/p_V$, the monochromatic polarized 850$\micron$ 
         intensity per unit V band polarization
         fraction for $\Tgr=18.5\K$ and $\Tgr=19.5\K$, for six
         porosities $\calP$.
         Green shaded region: $\pm10$\%
         range around the Planck results for translucent clouds
         \citep{Planck_int_results_xxi_2015,Planck_2018_XII}
         and
         diffuse clouds \citep{Planck_2018_XII}.
         Blue shaded region: 
         $\pm10\%$ range around results of
         PHS19 \citep{Panopoulou+Hensley+Skalidis+etal_2019}.
         Planck data have been color-corrected (see text).
         Red shaded region: axial ratios
         ruled out by $\langle\falign\rangle<0.7$ 
         (see Fig.\ \ref{fig:falign}).
         Filled symbols: cases that are compatible with observations
         (see text).
         \btdnote{f9a.pdf, f9b.pdf, f9c.pdf, f9d.pdf, f9e.pdf, f9f.pdf}
         }
\vspace*{-0.3cm}
\end{center}
\end{figure}
%%%%%%%%%%%%%%%%%%%%%%%%%%%%%%% end f9 %%%%%%%%%%%%%%%%%%%%%%%%%%%%%%%%%%

% sec 9
\section{\label{sec:polarized emission}
         Polarized Submm Emission}

At submm wavelengths, the electric dipole approximation is highly accurate.
The aligned grains required to explain the polarization of starlight
will generate submm emission, with polarized intensity
\beq
\frac{P_\nu}{\NH} =
V_{\rm align}\sin^2\gamma \left(\frac{\Cpolabs^\MPFA(\nu)}{V}\right)
B_\nu(T_{\rm gr})
~~,
\eeq
where $B_\nu(T)$ is the usual Planck function, and $T_{\rm gr}$ is the dust
temperature.
With $V_{\rm align}$ constrained by Eq.\ (\ref{eq:Valign})
so that astrodust material
reproduces the starlight polarization, this
becomes
\beq   % factor 1.82 = 1.823
\frac{P_\nu}{\NH} =
1.82\xtimes10^{-27}\cm^3\Ha^{-1} \sin^2\!\gamma
\times \frac{\Cpolabs^\MPFA(\nu)/V}{\Phisbg}
\xtimes B_\nu(\Tgr)
~~.
\eeq
The ratio $(\Cpolabs(\nu)/V)/\Phisbg$ (see lower panel
of Figure \ref{fig:Qpol850/Qabs850}) determines the ratio of polarized
emission per starlight polarization.  This ratio depends on both
axial ratio and porosity, and can therefore provide another
constraint on $b/a$ and $\calP$.
Prolate grains produce somewhat higher levels of polarized emission
(per starlight polarization) than do
oblate grains.

The fractional polarization of the astrodust emission is
\beq   
% factor 0.61 = 0.6077
p(\lambda) = 
\langle \falign^{\rm (Ad)} \rangle
\left[\frac{\Qpolabs^\MPFA(\lambda)}{\Qabs(\lambda)}\right]_{\rm Ad}
 \sin^2\!\gamma
= \frac{0.50(1-\calP)}{\Phisbg}\left[
        \frac{\Qpolabs^\MPFA(\lambda)}{\Qabs(\lambda)}\right]_{\rm Ad}
\sin^2\!\gamma
~,~~
\eeq
where Eq.\ (\ref{eq:falignsbg}) has been used.
The upper panel of
Figure \ref{fig:Qpol850/Qabs850} shows
$\Qpolabs^\MPFA/\Qabs$ at $850\micron$ as a function of axial ratio,
for different porosities.  As $\poro$ is increased, the polarizing
ability of the grain is reduced.

According to Figure \ref{fig:Qpol850/Qabs850} (lower panel),
$[\Cpolabs(850\micron)/V]/\Phisbg$ depends on
porosity $\poro$, but depends 
more strongly on grain shape, with higher values
for prolate shapes than for oblate shapes.
Prolate grains with $b/a=0.5$ and $\poro=0$ have $\Phi\approx 1.05$
(see Fig.\ \ref{fig:Phisbg vs ba}) and
$\Qpolext(850\micron)/\Qabs(850\micron)\approx 0.40$
(see Fig.\ \ref{fig:Qpol850/Qabs850}), giving
a submm polarization fraction $p_{850\mu{\rm m}}\approx 0.19\sin^2\gamma$
if the starlight polarization is $p_V \approx 0.130\sin^2\gamma E(B-V)/{\rm mag}$
(see Eq.\ \ref{eq:pmax propto E(B-V)}).
The polarized infrared intensity per unit starlight polarization is
(assuming $p_V\approx p_{\rm max}$)
\beqa   % Pi/pmax = 1.2344 -> 7.99MJy/sr * Cpol/Phi / (e^x-1)
\frac{P_\nu}{p_V}
&\,=\,&
\frac{\Piobs}{p_{\rm max}}
\times\frac{\Cpolabs^\MPFA(\nu)/V}{\Phisbg}\times B_\nu(\Tgr)
\\ \label{eq:P850/p_V}
\frac{P_\nu(850\micron)}{p_V}
&\approx& 7.99
\left(\frac{\Cpolabs^\MPFA(850\micron)/V}{\cm^{-1}}\right)
\frac{1}{\Phisbg} \left(\frac{1}{\exp(16.9\K/\Tgr)-1}\right) \MJy \sr^{-1}
~~~. 
\eeqa
Figure \ref{fig:P_nu/p_V vs ba} shows
$P_\nu(850\micron)/p_V$ as a function of axial ratio $b/a$, for
six different porosities.

For intermediate galactic latitudes dust temperatures have
been estimated to be $\Tgr=19.6\K$ \citep{Planck_int_results_xxii_2015}
and $\Tgr=19.4\K$ \citep{Planck_int_XLVII_2016}.
We show results for $\Tgr=19.5\K$, and also for $\Tgr=18.5\K$ to
show the sensitivity to the assumed dust temperature.

What is the actual value of $P_\nu(850\micron)/p_V$?
\citet{Planck_2018_XII} found 
$P_\nu(850\micron)/p_V=[5.42\pm0.05\MJy\sr^{-1}]/1.11$ for 1505 stars,
but a somewhat lower value,
$[5.0 \MJy\sr^{-1}]/1.11$, for low column densities.
Dust lying behind the star will raise 
$P_\nu/p_V$, but \citet{Planck_2018_XII}
argued that this bias was negligible.
A careful study 
of a small number of diffuse cloud 
sightlines using stars with Gaia distances
placing them beyond the dust obtained a lower value,
$P_\nu(850\micron)/p_V=[4.2 \pm 0.1 \MJy\sr^{-1}]/1.11$
\citep[][hereafter PHS19]{Panopoulou+Hensley+Skalidis+etal_2019}.
For all of these cases we apply a ``color correction'' factor $1.11$
to estimate monochromatic values at $353\GHz$
\citep[see discussion in][]{Hensley+Draine_2021a}.
These values are shown in Figure \ref{fig:P_nu/p_V vs ba}.
Why the PHS19 result differs from the \citet{Planck_2018_XII} result is not
clear.

Because of the uncertainty regarding the appropriate value of
$P_\nu(850\micron)/p_V$, we consider a dust model to be ``allowed'' if
$P_\nu(850\micron)/p_V$ falls anywhere in the range 
$3.44$ -- $5.37\MJy\sr^{-1}$ spanned by the \citet{Planck_2018_XII} 
and PHS19 results ($\pm10\%$).
A number of our astrodust models fall 
within this range.
The ones that also have
$\falign<0.7$ (i.e., outside the shaded red regions in 
Fig.\ \ref{fig:P_nu/p_V vs ba}) are shown with filled symbols.
Viable models include both oblate and prolate shapes.

%%%%%%%%%%%%%%%%%%%%%%%%%%%%%%%%%% f10 %%%%%%%%%%%%%%%%%%%%%%%%%%%%%%%%
\begin{figure}[ht]
\begin{center}
\includegraphics[angle=0,width=8.0cm,
                 clip=true,trim=0.5cm 5.0cm 0.5cm 2.5cm]%l b r t
{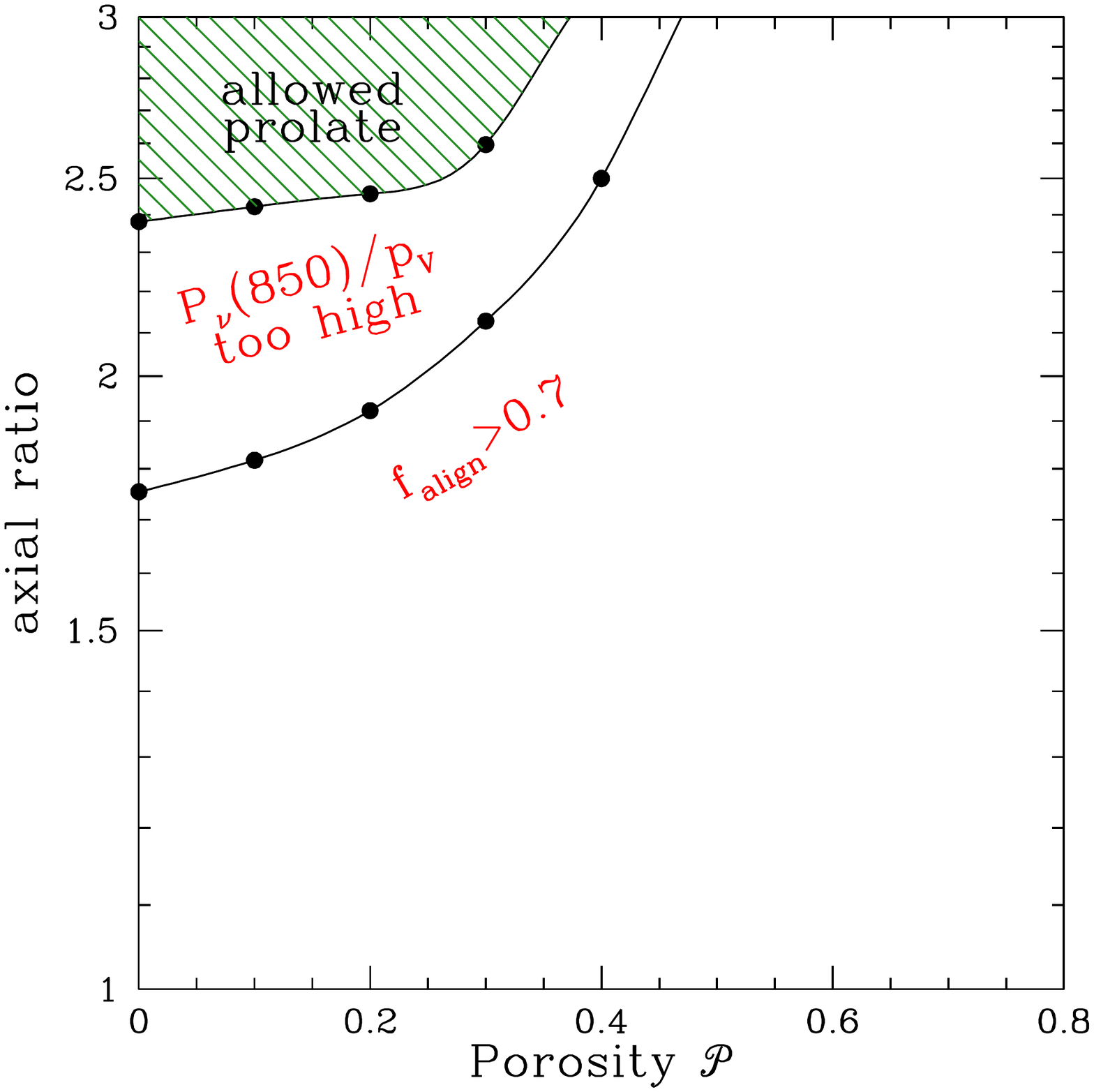}
\includegraphics[angle=0,width=8.0cm,
                 clip=true,trim=0.5cm 5.0cm 0.5cm 2.5cm]%l b r t
{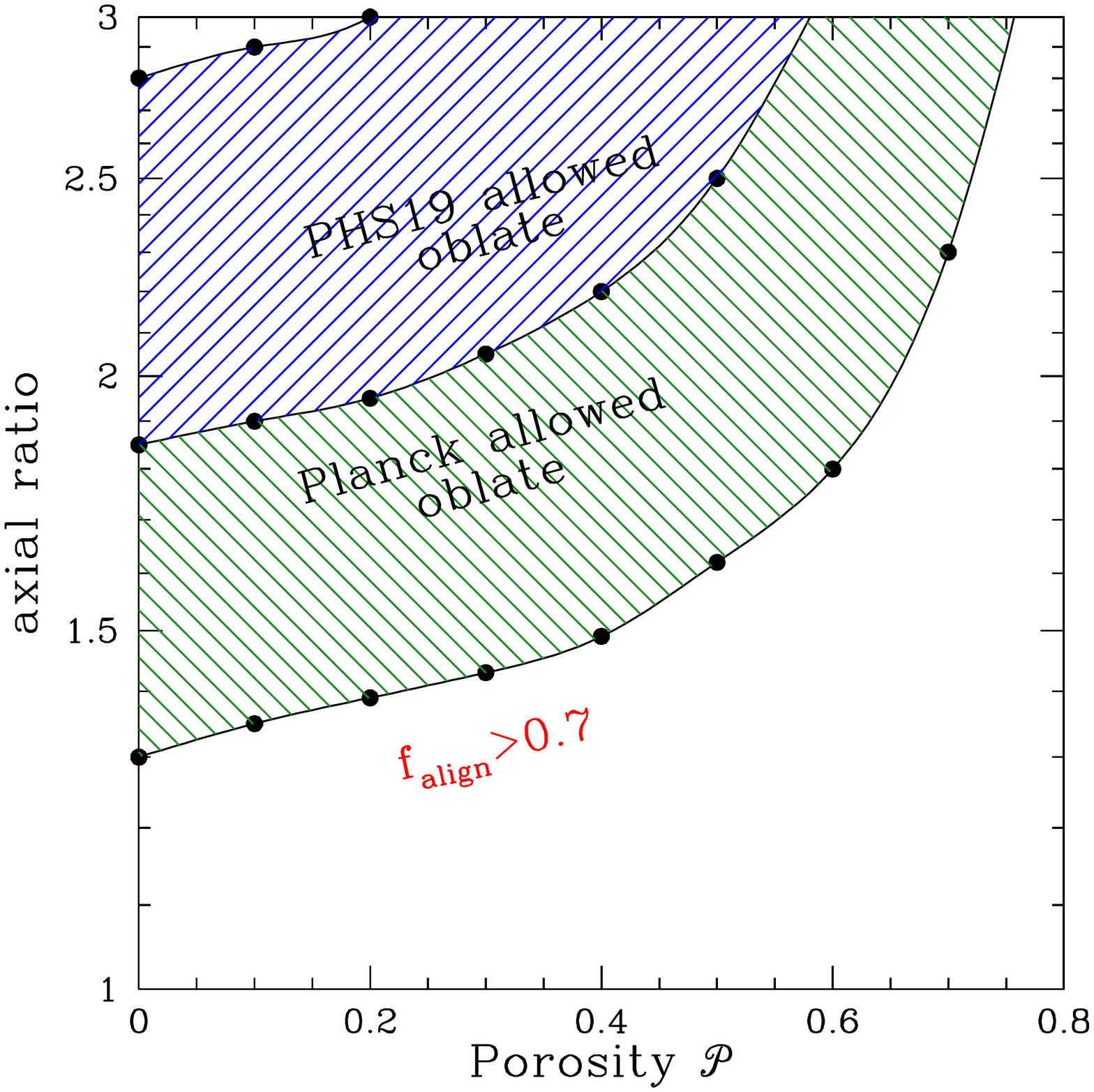}
\caption{\label{fig:allowed} \footnotesize
         Green shaded regions show allowed axial ratios for 
         prolate and oblate spheroids
         \btdnote{f10a.pdf,f10b.pdf}
         }
\vspace*{-0.3cm}
\end{center}
\end{figure}
%%%%%%%%%%%%%%%%%%%%%%%%%%%%%%%%% end f10 %%%%%%%%%%%%%%%%%%%%%%%%%%%%%%%%%%%

Figure \ref{fig:allowed} shows the allowed values of axial ratio and
porosity for prolate and oblate shapes.
For prolate spheroids 
with aspect ratio $a/b\leq3$, high porosities are excluded -- the
porosity $\poro<0.5$.
For oblate spheroids, somewhat larger porosities are allowed, but
the porosity is still limited to $\poro\ltsim 0.75$.  
%Even the high porosity
%cases are able to produce the values of $P_\nu(850\micron)/p_V$ found by
%PHS19,
%but not the higher values of $P_\nu(850\micron)/p_V$ found by
%\citet{Planck_2018_XII}.
%For spheroids with aspect ratios $\leq3$,
%porosities $\poro\gtsim0.75$ are excluded.

% sec 10
\section{\label{sec:10um pol}
         Silicate $10\mu$m Feature Polarization}

%%%%%%%%%%%%%%%%%%%%%%%%%%%%%%%% f11 %%%%%%%%%%%%%%%%%%%%%%%%%%%%%%%
\begin{figure}[ht]
\begin{center}
\includegraphics[angle=0,width=8.0cm,
                 clip=true,trim=0.5cm 5.0cm 0.5cm 2.5cm]%l b r t
{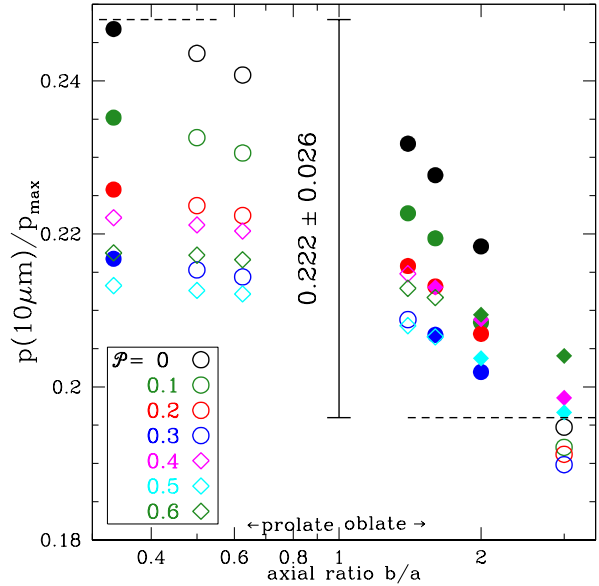}
\caption{\footnotesize \label{fig:p(10um)/pmax}
         Predicted ratio of 10$\micron$ polarization to peak
         optical polarization $p_{\rm max}$
         for astrodust spheroids with selected porosity $\calP$
         and selected axial ratios $b/a$.
         Filled symbols denote allowed cases 
         (see Figure \ref{fig:P_nu/p_V vs ba}).
         \btdnote{f11.pdf}
         }
\vspace*{-0.3cm}
\end{center}
\end{figure}
%%%%%%%%%%%%%%%%%%%%%%%%%%%%%%% end f11 %%%%%%%%%%%%%%%%%%%%%%%%%%%%%%%
If the astrodust grains are aligned, the
silicate absorption features at
10$\micron$ and $18\micron$ will be polarized.
Here we calculate the ratio of 
$p(10\micron)$ to the optical polarization $p_{\rm max}$.
Combining
\beq
p(10\micron) = \NH V_{\rm align}^{\rm (Ad)} \times
\left[\frac{\Cpolext^\MPFA(10\micron)}{V}\right]
\sin^2\!\gamma
\eeq
with Eq.\ (\ref{eq:pmax propto E(B-V)}) and
(\ref{eq:NH/E(B-V)}),
we obtain
\beq
\frac{p(10\micron)}{p_{\rm max}}
= \frac{\Piobs/p_{\rm max}}{\Phisbg}
\times
\left[\frac{\Cpolext^\MPFA(10\micron)}{V}\right]
~.~~
\eeq
Figure \ref{fig:p(10um)/pmax} shows the prediction for
$p(10\micron)/p_{\rm max}$ for different axial ratio $b/a$
and porosity $\calP$.  For the porosities $\calP\ltsim 0.5$ and
axial ratios that are consistent with the observed polarization of starlight
and polarized submm emission (see Fig.\ \ref{fig:allowed};
filled symbols in Figure \ref{fig:p(10um)/pmax}) 
we estimate that
\beq
\frac{p(10\micron)}{p_{\rm max}} \approx 0.222 \pm 0.026
\eeq
for $\Piobs/p_{\rm max} \approx 1.23\micron$
(see Table \ref{tab:tab1}).

To date there do not appear to be any published 
measurements of $p(10\micron)$ for sightlines
where the optical polarization $p_{\rm max}$ has also been measured.
The sightline to Cyg OB2-12 has 
$p_{\rm max}\approx p(0.43\micron)=0.0967\pm 0.0010$ 
\citep{Whittet+Martin+Hough+etal_1992}, and thus we predict 
$p(10\micron)\approx 0.021\pm0.003$ for Cyg OB2-12.

Measuring $p(10\micron)$ for Cyg OB2-12 
will be a valuable observational test for the hypothesis
that the starlight polarization, silicate absorption, and submm emission
arise from a single dominant grain type (``astrodust'').

% sec 11
\section{\label{sec:discussion}
         Discussion}

\citet[][hereafter GFV18]{Guillet+Fanciullo+Verstraete+etal_2018} 
developed models
using spheroids of amorphous silicate and of amorphous carbon
(a-C)
to reproduce both starlight polarization and 
polarized submm emission.
%GFV18 found that the polarization of starlight and the 
%polarized submm emission per
%starlight polarization 
%$P_\nu(850\micron)/p_V$ can be reproduced by prolate grains
%with a 2.5:1 axial ratio.
%We also find that prolate grains can reproduce the starlight polarization
%and polarized submm emission.
%
Both GVF18 and the present paper 
assume a population of ``large'' grains, plus a population of
PAH nanoparticles.
However GVF18 differs from the present paper in several respects.
For the large particles, GVF18 assumed silicate grains and 
a-C grains.
The silicate
grains were modeled using the ``astrosilicate'' dielectric function from
\citet{Weingartner+Draine_2001a}, with long-wavelength modifications from
\citet{Li+Draine_2001b}.
The a-C grains were modeled
using  optical constants of
``BE'' amorphous carbon 
from \citet{Zubko+Mennella+Colangeli+Bussoletti_1996}.
GVF18 also considered a model with astrosilicate and
a-C mixed in the same grains.

The present study assumes a population of ``astrodust''
grains, incorporating
both silicate and carbonaceous material,
with an effective dielectric function depending on assumed porosity, derived
as discussed by \citet{Draine+Hensley_2021a}.
These astrodust grains (plus a population of nanoparticles,
including PAHs) are able to reproduce the observed interstellar extinction
from the far-UV to $\sim$$30\micron$, as well as reproducing emission from
the mid-IR to the submm 
(B.S.\ Hensley \& B.T. Draine 2021, in prep.).
%\citep{Hensley+Draine_2021b}.

Despite the differences in assumptions of the two studies, 
similar conclusions are reached:
model D of GVF18 has aligned a-C grains that are prolate with 
axial ratio 3:1, and
aligned mixed silicate+a-C grains that are prolate with axial ratio 2.5:1.
Here we show that
the observed
starlight polarization integral $\Pi_\obs$ and polarized submm emission 
can be reproduced by ``astrodust'' spheroids with suitable
shape and porosity.
Viable cases (see Fig.\ \ref{fig:allowed}) include
2.5:1 prolate spheroids for $\poro<0.25$,
1.4:1 oblate spheroids with $\poro<0.25$,
or 2:1 oblate spheroids with $\poro < 0.65$.
%We have thus far 
The present study has only used an integral over the starlight polarization
as a constraint, allowing us to more thoroughly explore parameter space
(shape and porosity); 
future studies (Hensley \& Draine 2021, in preparation)
will employ models with detailed size distributions to reproduce the
wavelength dependence of both extinction and polarization.

One important conclusion of the present study is that 
extreme porosities are excluded.
High porosity grains are inefficient
polarizers, both for starlight polarization and submm polarization.
Models using spheroids with axial
ratios $\leq 3$ are able to reproduce the observed polarization of starlight
only for
porosity $\poro \ltsim 0.5$ for prolate shapes, or $\poro\ltsim0.75$
for oblate shapes (see Fig.\ \ref{fig:allowed}).
Some authors 
\citep[e.g.,][]{Mathis+Whiffen_1989,
Fogel+Leung_1998,
Min+Dominik+Hovenier+etal_2006,
Ormel+Min+Tielens+etal_2011,
Ysard+Jones+Demyk+etal_2018}
\btdnote{who else to cite?}
have proposed that interstellar grains may be highly porous, 
as the result of
coagulation processes.\footnote{E.g.,
   \citet{Ormel+Min+Tielens+etal_2011} considered aggregates with
   porosities as large as $\poro=0.9$,
   and \citet{Min+Dominik+Hovenier+etal_2006} considered
   some aggregates with $\poro>0.99$.}
However, we find here that the observed polarization of starlight, 
together with the strongly polarized thermal emission at submm wavelemgths, 
allows strong limits to be placed on the porosity.  
If the dust consists of either oblate or prolate spheroids with 
long:short axial ratios $\leq$\,2:1, 
the porosity of the dust cannot exceed $65\%$.
While low-velocity coagulation may tend to form ``fluffy'' structures, perhaps 
a combination of fragmentation and compression in higher velocity collisions
keeps the ``porosity'' of the dust population low.

The present work has been limited to a single family of very
simple grain shapes: spheroids.  
The combination of starlight polarization and submm polarization can
be reproduced with only certain axial ratios and porosities 
(Fig.\ \ref{fig:allowed}).
This in turn leads to a prediction for the polarization in the
10$\micron$ feature: $p(10\micron)/p_V = 0.222\pm0.026$.
How the ratio of submm polarization (grain in the Rayleigh limit) 
to optical polarization (grain size comparable to the wavelength) 
might vary for other grain geometries
is not yet known,
nor how this might impact predictions for
$p(10\micron)/p_V$.

It has been suggested
that the silicate grains may contain metallic
Fe inclusions, which could contribute magnetic dipole emission at
mm wavelengths \citep{Draine+Hensley_2013}.
If present, such Fe inclusions 
will affect their optical properties at all wavelengths.
The present paper has been limited to models where the silicate grains
do not contain metallic Fe inclusions.
We expect that the conclusions regarding grain shape and porosity will
not be substantially affected if the silicate grains contain Fe inclusions
with modest volume filling factors (e.g., $\ltsim$1\%), but this merits
future investigation if a signficant fraction of the Fe may be in
metallic inclusions.

Our understanding of grain shape can be expected to advance
as more measurements of starlight polarization become
available -- both the wavelength dependence for additional sightlines
\citep{Bagnulo+Cox+Cikota+etal_2017} as well as measurements of
starlight polarization $p_V$
for large numbers of stars
\citep[e.g., PASIPHAE,][]{Tassis+Ramaprakash+Readhead+etal_2018}
to allow determination of $P_\nu(850\micron)/p_V$ for many more sightlines.

The power of the present study has been limited by uncertainty
regarding the actual value of $P_\nu(850\micron)/p_V$.
The Planck result for $P_\nu(850\micron)/p_V$ for
translucent clouds is 29\% larger than the PHS19
value for selected sightlines; this range of values translates into
an enlarged domain of allowed values in the axial ratio-porosity plane.
Further studies of $P_\nu(850\micron)/p_V$, using starlight polarization
measurements on additional sightlines, will be valuable.  If
$P_\nu(850\micron)/p_V$ shows regional or environmental variations (beyond
what can be attributed to variations in grain temperature) this would be an
important clue toward understanding evolution of dust in the interstellar
medium.

Finally, we note that the profile of the $10\micron$ silicate feature in
polarization provides additional constraints on shape and porosity
\citep{Draine+Hensley_2021a}.  Unfortunately, existing spectropolarimetry
of this feature in the diffuse ISM is limited
\citep{Wright+Aitken+Smith+etal_2002}.
High signal-to-noise measurements of the $10\micron$ silicate polarization
profile would help constrain the porosity and shape of interstellar
grains.

% sec 12
\section{\label{sec:summary}
         Summary}

The principal results of this study are as follows:
\begin{enumerate}
\item The accuracy of the modified picket fence approximation (MPFA)
is tested at optical wavelengths.
The MPFA
provides an
adequate approximation to the 
orientationally-averaged
polarization profiles $\Cpolext(\lambda)$ for
spinning and precessing submicron grains.
At long wavelengths $\lambda\gg\aeff$ the
MPFA is highly accurate.

\item The {\it polarization efficiency integral} $\Phi$ 
(Eq.\ \ref{eq:PEI}) is introduced to measure
the effectiveness of grains for polarizing starlight.
We evaluate $\Phisbg(b/a,\aeff)$ for astrodust spheroids
(see Fig.\ \ref{fig:Phisbg vs ba}) with porosities from $\calP=0$
to $\calP=0.9$.

\item 
The fraction of the astrodust mass that is aligned,
$\langle\falign\rangle$, can be estimated from the
observed starlight polarization integral $\Piobs$ (Eq.\ \ref{eq:polint})
and the polarization efficiency integral $\Phisbg$,
without having to fit a size distribution of aligned grains to the
wavelength dependence of starlight polarization.

\item Assuming astrodust grains with spheroidal shapes, 
the limit $\langle\falign\rangle<0.70$, together with
the {\it starlight polarization integral} $\Piobs$ and the 
polarization efficiency integral $\Phisbg$, constrains
the aspect ratio of the dust grains producing the starlight polarization.
If the grains have low porosity,
then axial ratios $a/b>1.8$ are required if the grains are prolate 
spheroids, and axial ratios $b/a>1.4$ if the grains are oblate spheroids.
If the grains are substantially porous then
more extreme axial ratios would be required:
$a/b \gtsim 2.5$ or $b/a \gtsim 1.5$ for $\poro=0.4$ 
(see Fig.\ \ref{fig:falign}).

\item For spheroids with axial ratios $\leq3$,
the limit $\langle\falign\rangle<0.7$ and the observed
starlight polarization imply that extreme porosities $\poro\gtsim0.75$ are
excluded.
%\item When the observed submm polarization is used as an additional
%constraint, we find that interstellar dust particles, if approximated by
%spheroids, should be preferentially prolate, with axial ratios 
%$2\leq a/b\leq 3$, and porosities $\poro \ltsim 40\%$.

\item The ratio of polarized submm emission to starlight polarization
provides an additional constraint on porosity and grain shape.  We combine
this with the limit $\langle\falign\rangle<0.7$ to determine the
domain of allowed spheroid shapes and porosities (Fig.\ \ref{fig:allowed}).

\item We calculate the expected ratio of $10\micron$ polarization to
visual polarization $p_V$ if the grains can be approximated by spheroids.  
We predict 
$p(10\micron)/p_V=0.222\pm0.026$.
For Cyg OB2-12, we predict $p(10\micron)\approx (2.1\pm0.3)\%$. 
\end{enumerate}

\acknowledgments
This work was supported in part by NSF grants AST-1408723 and AST-1908123, and
carried out in part at the Jet Propulsion Laboratory, California
Institute of Technology, under a contract with the National Aeronautics
and Space Administration.
We thank
Francois Boulanger
and 
Vincent Guillet
for helpful discussions.
We thank Robert Lupton for availability of the SM package,
and the late Nicolai Voshchinnikov for generously 
making available the separation
of variables code {\tt hom6\_5q}.

\begin{appendix}
\section {\label{app:pfa} Orientational Averages}
 
Let $\langle ...\rangle$ denote averaging over orientation.
From Eq.\ (\ref{eq:ed 1},\ref{eq:ed 2}) we obtain the MPFA estimates
for the orientationally-averaged cross sections:
\beqa
\left\langle C_x \right\rangle^\MPFA
&=& 
\langle (\bahat\cdot\bznewhat)^2\rangle C_E(0) + 
\langle (\bahat\cdot\bxnewhat)^2\rangle C_E(90^\circ) + 
\langle (\bahat\cdot\bynewhat)^2\rangle C_H(90^\circ)
\\
\left\langle C_y\right\rangle^\MPFA
&=&
\langle (\bahat\cdot\bznewhat)^2\rangle C_E(0) + 
\langle (\bahat\cdot\bynewhat)^2\rangle C_E(90^\circ) + 
\langle (\bahat\cdot\bxnewhat)^2\rangle C_H(90^\circ)
\\ \label{eq:<Cx-Cy>MPFA}
\langle C_x-C_y \rangle^\MPFA
&=& 
\left[
\langle (\bahat\cdot\bxnewhat)^2\rangle -\langle (\bahat\cdot\bynewhat)^2\rangle
\right]
\left[C_E(90^\circ)-C_H(90^\circ)\right]
~~~.
\\
\langle C_x+C_y \rangle^\MPFA
&=& \nonumber 
2\left[1-\langle (\bahat\cdot\bxnewhat)^2\rangle-
        \langle (\bahat\cdot\bynewhat)^2\rangle
\right]C_E(0)
+
\\
&&
\hspace*{2em}
\left[
\langle (\bahat\cdot\bxnewhat)^2\rangle
+
\langle (\bahat\cdot\bynewhat)^2\rangle
\right]
\left[ C_E(90^\circ) + C_H(90^\circ)\right]
~~~.
\eeqa
These estimates for the orientationally-averaged cross sections
require knowledge of 
only three cross sections ($C_E(0), C_E(90^\circ), C_H(90^\circ)$) and
two moments of the distribution of orientations:
$\langle (\bahat\cdot\bxnewhat)^2\rangle$ and
$\langle (\bahat\cdot\bynewhat)^2\rangle$.
Randomly-oriented grains have
$\langle (\bahat\cdot\bxnewhat)^2\rangle =
\langle (\bahat\cdot\bynewhat)^2\rangle = \frac{1}{3}$.

\btdomit{``Perfect spinning alignment'' refers to
grains spinning with angular momentum $\bJ\parallel\bB_0$, 
with $\bJ$ aligned with the principal axis of largest moment of
inertia.
Oblate spheroids in perfect spinning alignment have 
$\langle (\bahat\cdot\bxnewhat)^2\rangle = \sin^2\gamma$,
$\langle (\bahat\cdot\bynewhat)^2\rangle = 0$.
Prolate spheroids in perfect spinning alignment have
$\langle (\bahat\cdot\bxnewhat)^2\rangle = \frac{1}{2}\cos^2\gamma$,
$\langle (\bahat\cdot\bynewhat)^2\rangle = \frac{1}{2}$.
}

If we continue to assume that the grains have their principal
axis of largest moment of inertia aligned with $\bJ$, but allow
$\bJ$ to make an angle $\psi$ with respect to $\bB_0$, then,
after averaging over rotation and precession of $\bJ$ around $\bB_0$,
and averaging over the distribution of $\psi$ values:
\beqa
{\rm oblate:} && \langle(\bahat\cdot\bznewhat)^2\rangle
                = \cos^2\gamma\langle\cos^2\psi\rangle + 
                  \frac{1}{2}\sin^2\gamma\langle\sin^2\psi\rangle
\\
              && \langle(\bahat\cdot\bxnewhat)^2\rangle 
                = \sin^2\gamma\langle\cos^2\psi\rangle + 
                  \frac{1}{2}\cos^2\gamma\langle\sin^2\psi\rangle
\\
              && \langle(\bahat\cdot\bynewhat)^2\rangle
                = \frac{1}{2}\langle\sin^2\psi\rangle
\\
              && \langle(\bahat\cdot\bxnewhat)^2-(\bahat\cdot\bynewhat)^2\rangle
                = \frac{3}{2}\sin^2\gamma
                  \left(\langle\cos^2\psi\rangle-\frac{1}{3}\right)
\\
{\rm prolate:}&& \langle(\bahat\cdot\bznewhat)^2\rangle
                = \frac{1}{4}\sin^2\gamma\left(1+\langle\cos^2\psi\rangle\right)
                  + \frac{1}{2}\cos^2\gamma\langle\sin^2\psi\rangle
\\
              && \langle(\bahat\cdot\bxnewhat)^2\rangle 
                = \frac{1}{4}\cos^2\gamma
                  \left(1+\langle\cos^2\psi\rangle\right) +
                  \frac{1}{2}\sin^2\gamma\langle\sin^2\psi\rangle
\\
              && \langle(\bahat\cdot\bynewhat)^2\rangle
                = \frac{1}{4}\left(1+\langle\cos^2\psi\rangle\right)
\\
              && \langle(\bahat\cdot\bxnewhat)^2-(\bahat\cdot\bynewhat)^2\rangle
                = \frac{3}{4}\sin^2\gamma
                  \left(\frac{1}{3}-\langle\cos^2\psi\rangle\right)
~~~.
\eeqa
\end{appendix}
\bibliography{/u/draine/work/libe/btdrefs}
\end{document}